# Single-molecule derivation of salt dependent base-pair free energies in DNA

Josep M. Huguet[a], Cristiano V. Bizarro[a,b], Núria Forns[a,b], Steven B. Smith[c], Carlos Bustamante[c,d], and Felix Ritort[a,b,1]

[a]Departament de Física Fonamental, Universitat de Barcelona, Diagonal 647, 08028 Barcelona, Spain. [b]CIBER-BBN de Bioingenieria, Biomateriales y Nanomedicina, Instituto de Sanidad Carlos III, Madrid, Spain. [c]Department of Physics, and [d]Department of Molecular and Cell Biology and Howard Hughes Medical Institute, University of California, Berkeley, CA 94720.

**Author contributions:** J.M.H. and F.R. designed research; J.M.H., C.V.B. and N.F. performed research; J.M.H., C.V.B., S.B.S. and F.R. analyzed data; and J.M.H., S.B.S, C.B. and F.R wrote the paper.

The authors declare no conflict of interest.

This article contains supporting information

[1] To whom Correspondence should be addressed.
Facultat de Física, UB
Diagonal 647, 08028 Barcelona, Spain
Phone: +34-934035869
Fax: +34-934021149
E-mail: fritort@gmail.com

# Abstract

Accurate knowledge of the thermodynamic properties of nucleic acids is crucial to predicting their structure and stability. To date most measurements of base-pair free energies in DNA are obtained in thermal denaturation experiments, which depend on several assumptions. Here we report measurements of the DNA base-pair free energies based on a simplified system, the mechanical unzipping of single DNA molecules. By combining experimental data with a physical model and an optimization algorithm for analysis, we measure the 10 unique nearest-neighbor base-pair free energies with 0.1 kcal $mol^{-1}$ precision over two orders of magnitude of monovalent salt concentration. We find an improved set of standard energy values compared with Unified Oligonucleotide energies and a unique set of 10 base-pair-specific salt-correction values. The latter are found to be strongest for AA/TT and weakest for CC/GG. Our new energy values and salt corrections improve predictions of DNA unzipping forces and are fully compatible with melting temperatures for oligos. The method should make it possible to obtain free energies, enthalpies and entropies in conditions not accessible by bulk methodologies.

The nearest-neighbor (NN) model (1-4) for DNA thermodynamics has been successfully applied to predict the free energy of formation of secondary structures in nucleic acids. The model estimates the free energy change to form a double helix from independent strands as a sum over all of resulting bp and adjacent-bp stacks, depending on the constituent four bases of the stack, by using 10 nearest-neighbor base-pair (NNBP) energies. These energies themselves contain contributions from stacking, hydrogen-bonding and electrostatic interactions as well as configurational entropy loss. Accurately predicting free energies has many applications in biological science: to predict self-assembled structures in DNA origami (5,6); achievement of high selectivity in the hybridization of synthetic DNAs (7); antigene targeting and siRNA design (8); characterization of translocating motion of enzymes that mechanically disrupt nucleic acids (9); prediction of non-native states (e.g. RNA misfolding) (10); and DNA guided crystallization of colloids (11).

Some of the most reliable estimates of the NNBP energies to date have been obtained from thermal denaturation studies of DNA oligos and polymers (2). Although early studies showed large discrepancies in the NNBP values, nowadays they are remarkably consistent among several groups. In these studies it is assumed that duplexes melt in a two-state fashion. However this assumption is not often the case and a discrepancy between the values obtained using oligomers vs. polymers remains a persistent problem that has been attributed to many factors such as the slow dissociation kinetics induced by a population of transient nondenatured intermediates that develop during thermal denaturation experiments (12). Single-molecule techniques (13) circumvent such problems by allowing one to control and monitor the denaturated state of a molecule along a full reaction coordinate. This paper reports measurements of the 10 NNBP energies in DNA by mechanically melting individual DNA molecules using an advanced optical tweezers apparatus. By measuring the force-distance curves (FDCs) we can determine the free energy of formation for the double helix. Previous studies have suggested using single-molecule force measurements to extract the NNBP energies in a wide variety of conditions (14). Here we show how, by combining developments in optical tweezers technology with refined data analysis, it is possible to determine free energy parameters with high precision (0.1 kcal/mol) in a wide variety of conditions, including salt concentration, pH and temperature. In particular, we have derived the salt corrections that apply to a wider range of salt (0.01-1 M NaCl) compared to the ranges (0.05-1 M NaCl) explored in thermal denaturation experiments (2).

## Results

Mechanical melting of DNA consists of pulling apart (unzipping) the two strands of a double-stranded DNA (dsDNA) molecule until the base pairs that hold the duplex together are disrupted and two single-stranded DNA (ssDNA) molecules are obtained. Such experiments reveal a FDC with a characteristic sawtooth pattern with force rips that are correlated with the DNA sequence (15-19). Features of this pattern are too coarse to distinguish individual base-pairs, but the energy of a particular type of bp stack can be inferred by its effect at many locations along the curve. To extract the NNBP energies high quality signal-to-noise measurements and reversible pulls are required. In previous studies of DNA unzipping either the length of the handles was too long (15,18,19) (permitting large thermal fluctuations in the molecular extension) or the experiments were performed at fast pulling speeds (16**)** and the unzipping/rezipping FDCs showed hysteresis (18) indicating that the process was not quasistatic. Here we demonstrate that by pulling DNA hairpins with extremely short handles at low pulling rates, one obtains FDCs that are essentially reversible (unzipping = rezipping). Besides, the slow pulling rate allows the system to visit states of higher energy at each fixed trap position. This permits to obtain an estimation of the equilibrium FDC (see *Materials and Methods*). These experimental FDCs can then be compared quantitatively with synthetic FDCs generated *in silico* by a physical model. To perform these tasks we have developed a miniaturized dual-beam optical tweezers apparatus (20, 21) (see *SI Appendix: Sections S1*, *S2*) and a curve-alignment algorithm to cancel instrument drift. The physical model involves an algorithm for searching the 10-dimensional space of possible NNBP energies that gives rapid and robust convergence to an optimum fit.

For the unzipping experiments, a molecular construct was synthesized starting from a 6,838 bp long DNA hairpin that was flanked by very short handles (29 bp) and a tetraloop (5'-ACTA-3') at its end (Fig. 1a and *SI Appendix: Section S3*). The molecule is tethered between an optically trapped bead and a bead at the tip of a pipette held by suction (Fig. 1b). DNA molecules are unzipped by moving the optical trap at low pulling speed (10 nm/s) and the reversible FDC is measured (Fig. 1c). Fig. 1d shows the unzipping FDCs of one molecule at various salt concentrations. The equilibrium FDC ($F_{eq}(x_{tot})$) describing the experiments can be obtained by computing the partition function of the system, *Z,* at a total distance $x_{tot}$ (see *Materials and Methods*),

$$
\begin{aligned}
Z(x_{tot}) &= \sum_n \exp\left(-\frac{G(x_{tot},n)}{k_B T}\right); \\
F_{eq}(x_{tot}) &= -k_B T \frac{\partial}{\partial x_{tot}} \ln Z(x_{tot}),
\end{aligned}
\qquad \textbf{[1]}
$$

where the sum in *Z* extends over all possible intermediate states (*n*). The free energy ($G(x_{tot},n)$) has three main contributions (see *SI Appendix: Fig.S1*). The first one is due to the stacking and hydrogen-bond energies of the bases. while the second one comes from the elastic contribution of the released ssDNA during unzipping. The third one is the parabolic potential that allows us to progressively unzip the molecular construct and access to any particular region of the sequence. When dsDNA is melted mechanically, the two product strands of the reaction are produced under a tension around 15 pN, whereas a standard melting reaction produces ssDNA strands under zero tension. Therefore to achieve the same final state as a standard reaction, one needs to relax the separate strands and measure the energy that is returned. Strands of ssDNA exhibit a complex form of entropic elasticity which can be modeled numerically (22). A set of empirical fits to ssDNA elasticity was developed based on pulling experiments done with a 3 kb piece of ssDNA (Fig. 1e and *SI Appendix: Section S4*). We have found that the worm-like-chain (WLC) model correctly fits the ssDNA elastic response for salt concentrations below 100 mM NaCl, whereas above this value FDCs develop a plateau at low forces related to the formation of secondary structures (23). Above the force plateau the freely-jointed chain (FJC) model fits data better (*SI Appendix: Section S5*). In order to extract the NNBP free-energy changes for a molecule that adopts a (hypothetical) non-self-interacting melted state it is best to exclude the formation of partial secondary structures in the initial and final states. This exclusion can be achieved by interpolating a FJC fit between the high force region (e.g. above 15 pN) and the point of zero force, i.e. zero molecular extension. Here we are assuming that the ssDNA has no (secondary) structure at high salt concentrations. Our measurements give the free-energy difference between two ideal ssDNA complementary strands and the hybridized dsDNA duplex. This assumption holds at low salt concentrations and is an approximation at high salts. Discounting secondary structures in the unzipped "coil" state is an improvement over previous bulk methods where they could not detect such structures and these structures violate the two-state hypothesis (12). We have checked that the obtained elastic properties of the ssDNA match the last part of the unzipping FDC, when the molecule is fully extended (Fig. 2a,b, magenta curve).

When we include the effects of elasticity and calculate the FDC by using the NNBP energies provided by SantaLucia (2) (i.e. the Unified Oligonucleotide (UO) energies) and currently used by Mfold (24-26), we observe a qualitative agreement with the experimental FDC (Fig. 2a,b, black and blue curves). Nevertheless, slight but systematic deviations between both curves are observed, particularly when considering the effect of salt concentration. Although the mean unzipping force predicted by the UO energies shows a logarithmic dependence with salt concentration (Fig. 2c, green curve) it overestimates the measured values by nearly 1 pN at low salt. These differences indicate a slight error in the NNPB UO energies. According to Eq. **1** an average 8% correction in the NNBP energies introduces a difference of 1 pN in the mean unzipping force. The best values for the NNBP energies ($\varepsilon_i \; i=1,...,10$) can be inferred by minimizing the mean squared error between the experimental and theoretical FDCs:

$$E(\varepsilon_1,...,\varepsilon_{10},\varepsilon_{loop}) = \frac{1}{N}\sum_{i=0}^{N}\left(F_i^{exp} - F_i^{the}(\varepsilon_1,..,\varepsilon_{10},\varepsilon_{loop})\right)^2 \quad \textbf{[2]}$$

where $E(\varepsilon_1,...,\varepsilon_{10},\varepsilon_{loop})$ is the total error, $N$ is the number of experimental points in the FDC, and $F_i^{exp}$ is the experimental equilibrium force, averaged with a low-pass filter of 1 Hz to avoid hopping artifacts. In order to extract the best values for the $\varepsilon_i$, Eq. **2** is minimized in an 11-dimensional space using a Monte Carlo based approach (see *SI Appendix: Section S6*). Moreover, in order to correct for the position drift in the optical tweezers, a shift function in the model was introduced that locally shifts the position of the trap along the FDC. The shift function leads to an improved match between theoretical and experimental FDCs without affecting the NNBP energy values (*SI Appendix: Section S7*). Our best-fit energy parameters reduce the error (i.e. give improved agreement) between the measured FDC and the theoretical prediction (Fig. 2a and b, red and black curves). By minimizing the error function Eq. **2** and estimating the NNBP energies for many individual molecules, we can obtain error limits on the free-energy values. In Fig. 2d and e we plot the average value and the standard error of the NNBP obtained for the 6.8 kb sequence and the UO prediction for the NNBP energies at 10 mM and 1 M NaCl. It is interesting that some of the new values are in good agreement with the results computed by SantaLucia (2) (e.g. CA/GT and AT/TA motifs) while others differ significantly (e.g. AA/TT and GA/CT at 10 mM NaCl and AC/TG and CC/GG at 1 M NaCl). According to the UO salt correction, the NNBP energies are extrapolated homogenously (i.e. the same salt correction is taken for all base-stack combinations) from standard salt conditions (1 M NaCl) down to lower salt concentrations (e.g., 50 mM). However, such correction does not predict the observed unzipping force at low salt, especially for certain NNBP such as AA/TT or GA/CT. This discrepancy is somehow expected at the lowest salt regime (10-50 mM NaCl) since the UO salt correction applies between 0.05-1 M NaCl. Nevertheless, we also observe discrepancies in the mean unzipping force at salt concentrations above 100 mM NaCl. A heterogeneous (sequence specific) salt correction could provide consistent results with the experiment. Such deviations are not unexpected, given the differences in solvation between specific nucleotides and salt ions (27,28), however the effect has never been quantified in the context of polymeric DNA. With this goal in mind, we applied our fitting algorithm to extract NNBP energies for data taken at many salt concentrations (Fig. 3, red points). As a further check, we repeated the experiments at 500 mM and 1 M NaCl by using a different test sequence from the other end of lambda (see *SI Appendix: Section S3*). Thus we found compatible NNBP energies between the two test molecules (Fig. 2e and Fig. 3, blue dots).

The UO model uses a non-sequence-specific salt concentration correction for the different NNBP energies given by:

$$\varepsilon_i\left(\left[\mathrm{Mon}^+\right]\right)=\varepsilon_i^0-m\cdot\ln\left(\left[\mathrm{Mon}^+\right]\right), \qquad \textbf{[3]}$$

where $\varepsilon_i([\mathrm{Mon}^+])$ is the energy of formation of the $i$th NNBP ($i$=1,...,10) at monovalent salt concentration $[\mathrm{Mon}^+]$ (expressed in M units), $\varepsilon^0_i$ is the NNBP energy at 298 K, 1 M monovalent salt and $m$ is the non-specific prefactor equal to $m$=0.110 kcal/mol (2,25) at 298 K (Fig. 3, green lines). The UO model assumes that only the entropy (and not the enthalpy) depends on the salt and this dependence is uniform at all temperatures meaning that $m(T)/T$ is a constant. Therefore, the correction of salt for the free energy changes depends on the temperature according to $m(T)=\frac{T}{T_0}m(T_0)$ where $T_0$ is a reference temperature (see *SI Appendix: Section S8*). To make a heterogeneous salt correction, one needs only to define 10 sequence-specific prefactors $m_i$ to be used with the same logarithmic dependence as shown in Eq. **3**. Thus we fit all NNBP energies using NNBP-dependent parameters $m_i$ ($i$ = 1,..10, *loop*) and $\varepsilon_i^0$. Such a fit is shown in Fig. 3 (black lines) and best values for $m_i$ are listed in Table 1. There we observe that the salt dependence of some NNBP parameters is well described by the UO nonspecific correction (e.g., AT/TA and CA/GT) but most of them are better fit with some correction in parameters $\varepsilon^0_i$ and $m$ (e.g., AA/TT, AC/TG, AG/TC). We have noticed that NNBP purine-purine or pyrimidine-pyrimidine combinations (5’-YY-3’ or 5’-RR-3’, i.e., AA/TT, AG/TC, CC/GG, GA/CT) differ most from the UO homogeneous salt correction than mixed purine-pyrimidine combinations (5’-RY-3’ and 5’-YR-3’). A difference between these combinations can be observed in how charges (e.g., hydrogen-bond acceptor and donor groups) are distributed along the major groove of the double helix. The latter have charged groups that tend to be uniformly distributed between the two strands along the major groove, whereas the former have donor and acceptor groups unevenly distributed between the two strands. The specific salt correction found in our measurements could be consequence of how monovalent cations bind the two strands along the major groove. There are precedents to such results: Sugimoto and collaborators (29) have reported that cation binding is correlated to duplex stability. Computer simulations have identified acceptor groups in guanine (N7, O6) and adenine (N7) as preferential cation binding sites (30). Our experimentally determined specific salt corrections might be interpreted as a corroboration of such hypothesis.

Finally, we wished to check how well our unique free-energy values work to predict the melting temperature of oligonucleotides under various salt conditions. Fortunately, there are several published studies giving accurate experimental values. Although $T_M$ is not a robust estimator to compare the melting and unzipping experiments, this is the most reliable experimental observable from melting data with which we can compare our results. For non-self-complementary oligos, the melting temperatures can be estimated from the following expression (2)

$$T_M = \frac{\Delta H^0}{\Delta S^0 + \sum_i \frac{m_i(T)}{T} \ln\left[Mon^+\right] + R\ln[C_T/4]} \qquad \textbf{[4]}$$

where $\Delta H^0$ and $\Delta S^0$ are the oligo enthalpy and entropy that are assumed to be temperature independent, $m_i$ are taken at $T = 298$ K and $C_T$ is the total single-stranded concentration of the oligo. In order to compare UO free energies with the unique values, we have recalculated the melting temperatures of 92 oligos at five different salt conditions and compared our results with melting data taken from Owczarzy et al. (31). By taking the UO values for the NNBP free energies, enthalpies and entropies (with the corresponding initiation factor for each oligo, see *SI Appendix: Section S8 and Table S1*) at standard conditions (1 M NaCl) but using the heterogeneous salt correction, the error committed in the extrapolation at lower salts is found to be below 2 ºC, which is similar to the error of the UO model (Fig. 4a). This fact reveals that the average salt correction for all NNBP, $m = \frac{1}{16}\sum_{i=0}^{16} m_i = 0.104$, is equivalent to the homogeneous UO correction ($m$ = 0.110) at 25 ºC. However, a closer inspection shows that a heterogeneous salt correction does a better job in predicting melting temperatures than the homogeneous one (Fig. 4b) for oligos longer than 15 bp. We have fixed the 10 values for the NNBP free energies $\varepsilon_i$ and the 10 parameters $m_i$ as given by our measurements and determined the 10 enthalpies $\Delta h_i$ that minimize the error function $\chi^2 = \frac{1}{N}\sum_i^N \left(T_i^{exp} - T_i^{pred}\right)^2$, where $i$ runs over all ($N = 460$) oligos and salt conditions shown in ref. 31 (see *SI Appendix: Section S9*). Here $T_i^{exp}$ is the melting temperature experimentally measured in ref. 31 and $T_i^{pred}$ is the melting temperature predicted by Eq. **4**. For oligos longer than 15 bp, the UO parameters with homogeneous salt corrections give $\chi^2$ = 2.37 (corresponding to 1.5 °C average error) whereas the optimal enthalpies (Table 2) give a lower error, $\chi^2$=1.74 (1.3 °C average error). For oligos of 15 bp or shorter our best values underestimate melting temperatures by 2-4 °C. Finally we note that a moderate increase of the NNBP values by 0.15 kcal/mol (i.e., slightly beyond the standard error given for the NNBP values) makes the standard deviation error for temperature melting prediction to go from ≅2 °C up to 5-6 °C.

## Discussion

Why do our free-energy numbers predict fairly well melting temperatures of oligos longer than 15 bp but do worse for shorter ones? Discrepancies between predicted and measured melting temperature for short oligos have been already reported in bulk measurements (32) and attributed to differences in analytical methods used to extract melting temperatures. Another possible explanation is that short oligos (≤ 15 bp) might not have the double helix perfectly formed and the formation energies involved in the duplex are slightly different from the energies for longer sequences. Although we lack a conclusive answer to this question, it is worth underlining that UO free-energy values are obtained in order to correctly predict the melting temperatures for all oligo lengths. This constraint might lead to error compensation between the melting temperature datasets corresponding to short and long oligos. Let us stress that with increasing length, the $T_M$ prediction is more tolerant of errors in the details of the NNBP energies, where sequence effects are averaged out. In addition, deviations from the bimolecular model (*SI Appendix: Section S8, Eq. 9*) arise for sequences with $n \geq 20$, as their melting process begins to shift toward pseudomonomolecular behavior (33). Still, our predicted melting temperatures for oligos with $n \geq 20$ agree well for the sequences reported in ref. 31.

We have performed single-molecule force unzipping experiments to extract DNA base pair free energies at various salt concentrations finding heterogeneous salt corrections. What is the origin of specific salt corrections? As previously said, this specificity might be consequence of how donor and acceptor groups distribute between the two strands along the major groove of the helix. However, there is an alternative interpretation based on the sequence dependence of ssDNA elasticity. Previous studies (34) have suggested a conformational transition of the sugar pucker in ssDNA that goes from the A-form (C3'-endo) at low forces to the B-form (C2'-endo) at high forces. A related phenomenon has been reported in recent stretching studies of homopolymeric RNA sequences (35) that reveal sequence dependent base-stacking effects. Based on our experimental data we cannot discard such interpretation. An exhaustive research of the elastic response of different homopolymeric ssDNA sequences (spanning different combinations of stacked bases) could shed light into this question.

The new values for the NNBP energies reported here are compatible with force unzipping experiments and improve melting temperature prediction as compared to the UO prediction for oligos longer than 15 bp. Although melting and unzipping experiments are based on disruption processes triggered by different external agents (temperature and force respectively), the agreement between the NNBP energies obtained is remarkable. Our work shows that using very different experimental systems the NN model can provide remarkably consistent results. The new NNBP energies predict both the unzipping experiments and the melting temperatures of oligos fairly well at low temperatures and low salt concentration. In these experimental conditions, the unzipping experiments provide an alternative determination of the NNBP parameters that seems to work better than the UO parameters. One important advantage over optical melting experiments is that the folding/unfolding transition does not need to be two-state. Besides, instead of several short oligos of different sequence, one long molecule is sufficient to infer the NNBP energies. The main limitation of our method is the accurate determination of the elastic response of the ssDNA. A 10% error in the estimation of the persistence or Kuhn length of ssDNA induces a similar error in the prediction of the NNBP energies. Moreover, the bimolecular initiation factors cannot be determined with our methodology. This approach can be extended to extract free energies, entropies and enthalpies in DNA and RNA structures under different solvent and salt conditions. To estimate NNBP entropies and enthalpies we should have to perform experiments at different temperatures. At present this cannot be achieved with our experimental setup because the changes of temperature dramatically affect the optics of the instrument. Temperature variations introduce undesirable drift effects that compromise the resolution of the measurements. The method can be also applied to extract free energies of other structural motifs in DNAs (e.g., sequence dependent loops, bulges, mismatches, and junctions). Most important, force methods make it possible to extract free energies in conditions not accessible to bulk methods. One example is unzipping dsRNA in the presence of magnesium where free-energy prediction is not possible from melting experiments because RNA hydrolyzes below the melting temperature. Another example is binding free energies of DNAs and RNAs bound to proteins where the proteins denaturalize below the dissociation melting transition. Although binding/unbinding FDCs could possibly not be reversible, nonequilibrium free-energy methods could be used to extract the binding free energy. This method could be also useful in cases where molecular aggregation and other collective effects in bulk preclude accurate free-energy measurements. Finally, the NN model (with adjusted parameters) has predicted the experimental FDC data very well, but we have only explored two specific sequences from lambda. By using other dsDNA molecules, we could search for long-range context effects (e.g., second-nearest and third-nearest interactions) in precisely those places along the sequence where the present model might be seen to fail. Because the mechanical unzipping method can be performed in many sequences it may be time to test the applicability of the next-nearest-neighbor model.

Our work establishes a unique methodology to obtain thermodynamic information from single-molecule experiments. Future experiments will address the question of the sequence-specific salt effects in the ssDNA. A different expansion of our work will determine the NNBP enthalpies by performing experiments at different temperatures. The use of stiffer optical traps, which makes the partition function more localized, will provide more precise energy measurements. Stiffer optical traps will be a starting point to observe second NN effects.

## Materials and Methods

The FDC is calculated by using a mesoscopic model that describes separately each component of the experimental setup (36): the bead in the optical trap, the handles, the released ssDNA, and the dsDNA hairpin (see *SI Appendix: Fig. S1*). The potential energy of the bead in the optical trap is described by a harmonic potential which is determined by the stiffness of the trap $E_b(x_b) = \frac{1}{2} k x_b^2$, where $k$ is the trap stiffness and $x_b$ is the elongation of the bead from the center of the trap. We use the NN model to describe the free energy of formation of the DNA duplex (2). Since DNA has four types of bases (adenine, guanine, cytosine and thymine), the NN model should have 16 different parameters. However, due to symmetry considerations, there are only 10 independent parameters. The free energy required to open $n$ bps is given by the sum of the free energies required to open each consecutive NN pair $G_{DNA}(n) = \sum_{i=0}^{n} \varepsilon_i$, where $G_{DNA}(n)$ is the free energy of the hairpin when $n$ bps are disrupted and $\varepsilon_i$ is the free energy required to disrupt the bp $i$. Therefore, the free energy of formation of the duplex depends on the sequence of bp. The NN model assumes that only local interactions between base pairs are relevant. It is also assumed that each interaction (composed of hydrogen bonding, stacking and entropy loss) can be described by one single free-energy value. An extra free-energy contribution is included in the model to account for the disruption of the end loop (see *SI Appendix: Section S10*). Elastic models for polymers are used to describe the elasticity of the handles and the ssDNA released during the unzipping process. The handles are dsDNA and they are modeled using the force vs. extension curve of a WLC $F(x_h) = \frac{k_B T}{4 l_p}\left(\left(1 - \frac{x_h}{L_0}\right)^{-2} - 1 + 4\frac{x_h}{L_0}\right)$, where $k_B$ is the Boltzmann constant and $T$ is the temperature, $l_p$ is the persistence length and $L_0$ is the contour length. The elastic free energy of the handles is obtained by integrating the previous expression. The ssDNA is modeled using either a WLC or a FJC model, depending on the salt concentration of the experiment (see *SI Appendix: Section S5*). In the case of the FJC model, the following equation gives the extension vs. force curve, $x_s(F) = L_0\left(\coth\left(\frac{bF}{k_BT}\right) - \frac{k_BT}{bF}\right)$, where $b$ is the Kuhn length. Again, the elastic free energy of the ssDNA is obtained by integrating the force vs. molecular extension curve. The parameters that define the elastic response of the handles are taken from the literature (34): $l_p = 50$ nm and $L_0 = 9.86$ nm (= 0.34 nm/bp × 29 bp). The total free energy of the total system is given by the sum of all free-energy contributions $G(x_{tot},n) = E_b(x_b) + 2G_h(x_h) + 2G_s(x_s,n) + G_{DNA}(n)$. The total distance of the system is given by the sum of all extensions corresponding to the different elements $x_{tot} = x_b + 2x_h + 2x_s$ (see *SI Appendix: Fig. S1*). The total free energy of the system is completely determined by $x_{tot}$ and $n$. The equilibrium FDC can be numerically calculated via the partition function defined by Eq. **1** and where we only include sequential configurations (*SI Appendix: Section S11*). *SI Appendix: Section S12* describes the thermodynamic process of unzipping.

**ACKNOWLEDGMENTS.** We thank T. Betz, M. Orozco, J. Subirana and I. Tinoco Jr. for useful comments and a careful reading of the manuscript. J.M.H was supported by the Spanish Research Council in Spain. F.R. is supported by Grants FIS2007-3454 and Human Frontier Science Program (HFSP) (RGP55-2008).

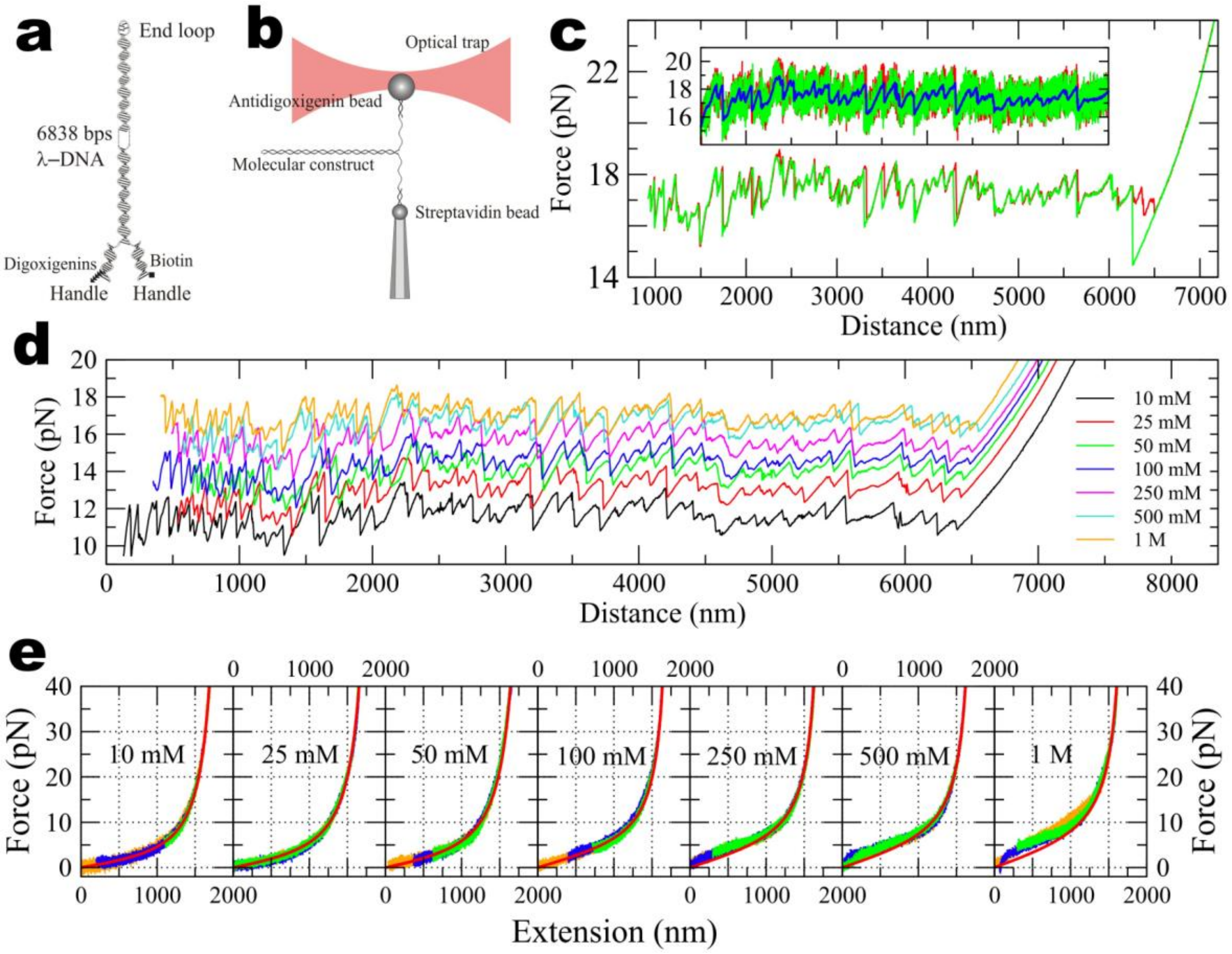

**Figure 1.** Experimental setup and results. (**a**) Molecular construct. A sequence of 6,838 bp obtained from λ-DNA is ligated to a tetraloop and two short handles of 29 bp each. **(b)** Experimental setup. The molecular construct is attached to two beads. The unzipping experiment is performed by moving the optical trap relative to the pipette. (**c)** Unfolfing (red) and refolding (green) curves filtered with a running average filter of 1 Hz bandwidth. There is always some hysteresis in the last rip. Pulling and relaxing are almost identical in the rest of the curve. Inset shows raw data (same color as before, blue curve is the average at bandwidth 1 Hz). (**d)** FDCs at various monovalent salt concentrations. (**e)** Elastic response of a 3 kb ssDNA molecule at various salts. Raw data of three molecules are shown (orange, green and blue curves). Red curve shows the best fit to the elastic model.

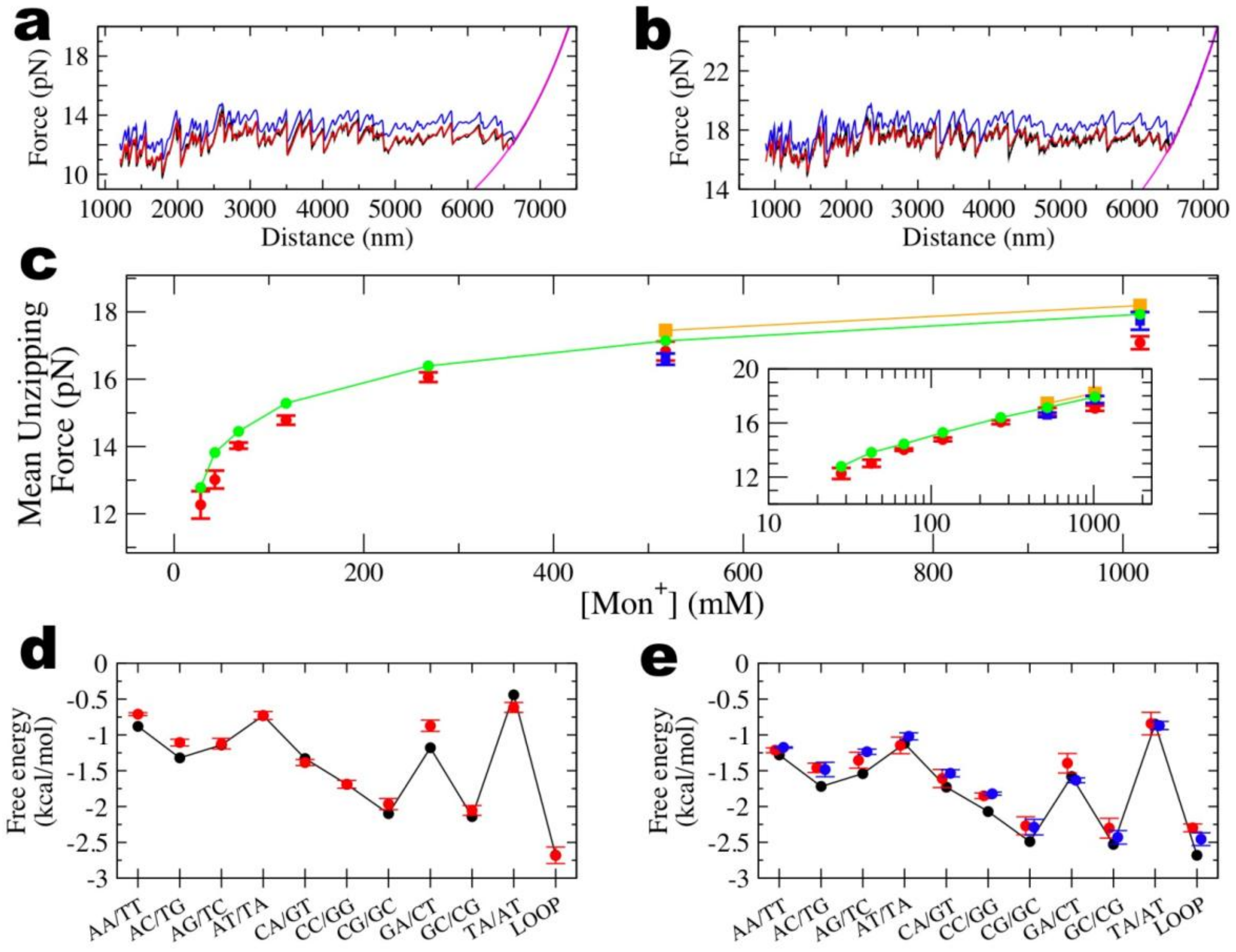

**Figure 2.** Salt dependencies. **(a,b)** FDCs for the 6.8 kb sequence at 10 mM NaCl (**a**) and 1 M NaCl (**b**). Black curve, experimental measurements; blue curve, UO prediction; red curve, our fit; magenta curve, elastic response of the fully unzipped molecule. The theoretical FDC is calculated in equilibrium, which assumes that the bandwidth is 0 Hz and the experimental data is filtered at bandwidth 1 Hz. If data is filtered at higher frequencies (> 1 Hz), hopping between states is observed and the experimental FDC does not compare well with the theoretical FDC at equilibrium. If data is filtered at lower frequencies (< 1 Hz), the force rips are smoothed and hopping transitions are averaged out. (**c**) Dependence of mean unzipping force with salt concentration. Red points, experimental measurements for the 6.8 kb sequence; green curve, UO prediction for the 6.8 kb sequence; blue points, experimental measurements for the 2.2 kb sequence; orange curve, UO prediction for the 2.2 kb sequence. **(d,e)** NNBP energies

and comparison with UO values at 10 mM NaCl (**d**) and 1M NaCl (**e**). The following notation is used for NNBP: AG/TC denotes 5´-AG-3´ paired with 5´-CT-3´. Black points, UO values; red points, values for the 6.8 kb molecule; blue points, values for the 2.2 kb molecule. The values for the 6.8 kb and the 2.2 kb molecules have been obtained after averaging over six molecules. Error bars are determined from the standard error among different molecules.

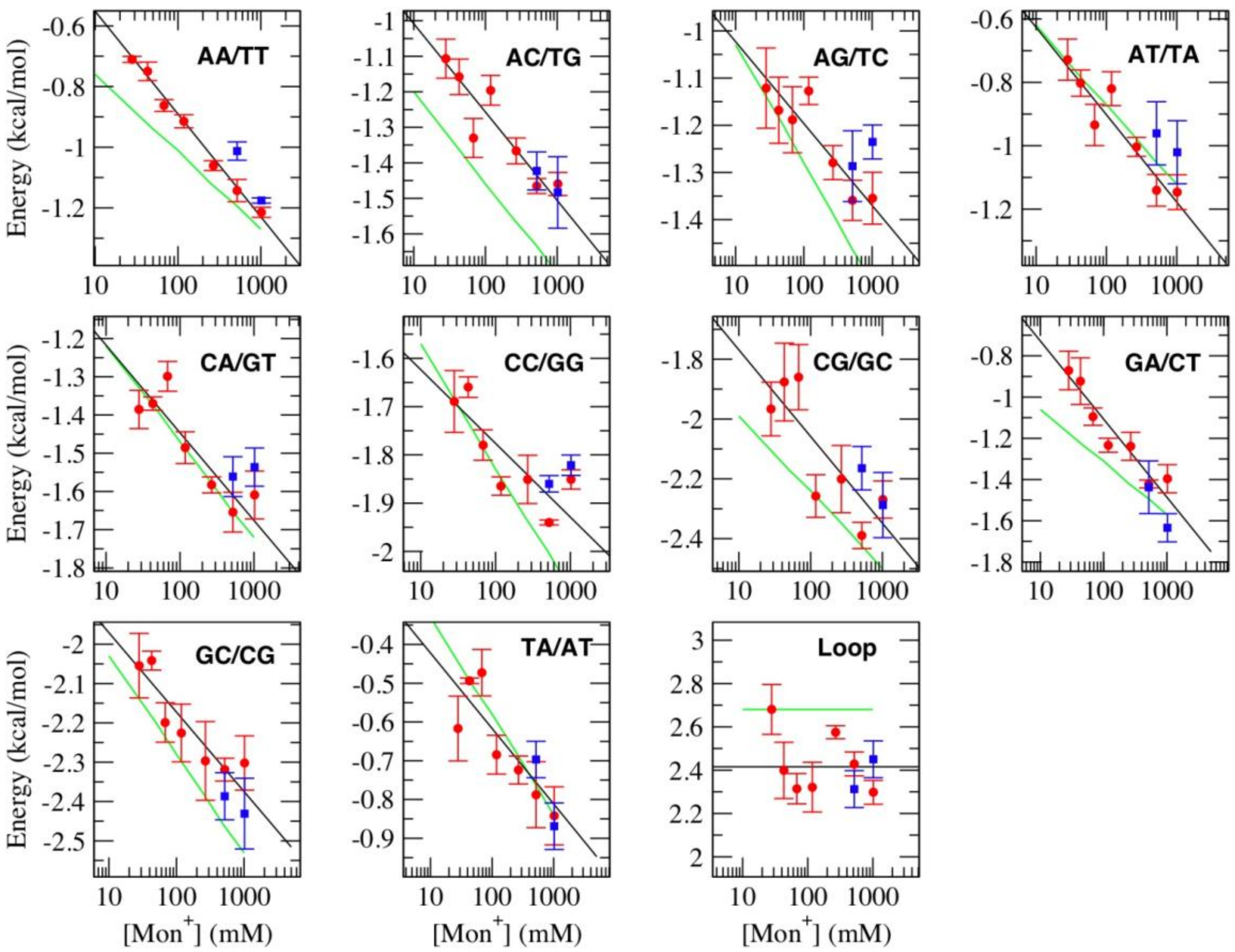

**Fig. 3.** Salt corrections of the NNBP energies. Figure shows the energy of all different NNBP parameters. Red (blue) points are the experimental results for the 6.8 kb (2.2 kb) sequence; green curve, UO nonspecific salt correction; black curve, fit to Eq. **3** with adjustable parameters $m_i$ ($i$=1,..10,$loop$) and $\varepsilon^0_i$.

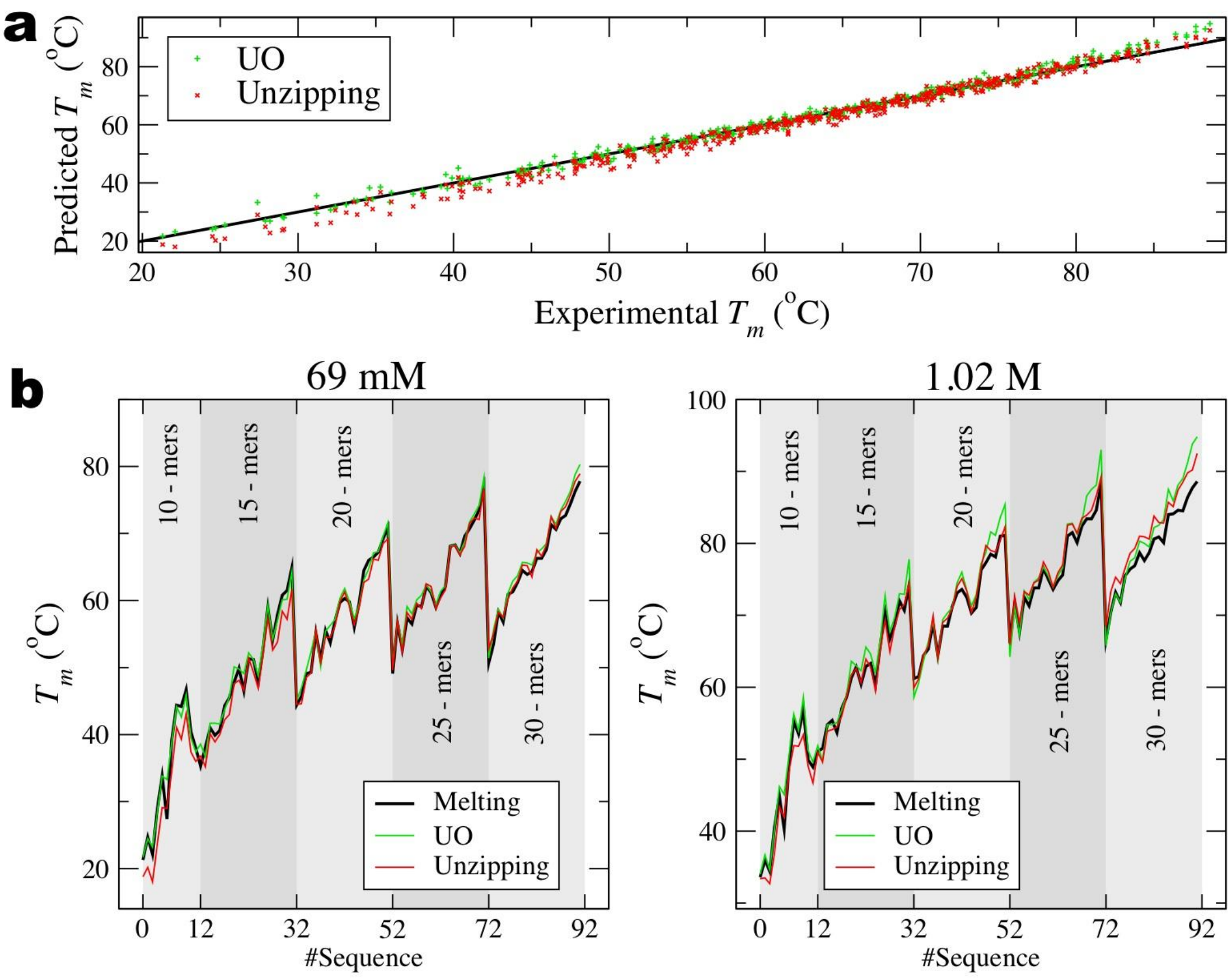

**Fig. 4.** Melting temperatures prediction. Comparison with melting temperatures for the 92 oligos ranging from 10-30 bp reported in ref. 31 (data reported in *SI Appendix: Table S2*). **(a)** Predicted vs. experimentally measured melting temperatures at five salt conditions ([$Na^+$] = 69, 119, 220, 621, 1,020 mM). The values obtained from unzipping have less error at higher temperatures (corresponding to longer oligos). **(b)** Prediction at 69 mM (left) and 1.02 M NaCl (right). Black lines are the experimentally measured melting temperatures, green line is the UO prediction and red line our prediction from unzipping data.

| NNBP | 6 kb | 2kb | Best | UO | $\varepsilon^o_i$ | $m_i$ |
|---|---|---|---|---|---|---|
| AA/TT | -1.21 (0.02) | -1.18 (0.01) | -1.20 (0.02) | -1.27 | -1.23 (0.01) | 0.145 (0.006) |
| AC/TG | -1.46 (0.04) | -1.48 (0.10) | **-1.47 (0.10)** | -1.71 | -1.49 (0.05) | 0.10 (0.02) |
| AG/TC | -1.35 (0.07) | -1.24 (0.04) | **-1.30 (0.07)** | -1.53 | -1.36 (0.03) | 0.070 (0.014) |
| AT/TA | -1.15 (0.06) | -1.02 (0.05) | -1.09 (0.08) | -1.12 | -1.17 (0.04) | 0.12 (0.02) |
| CA/GT | -1.61 (0.07) | -1.54 (0.05) | -1.58 (0.08) | -1.72 | -1.66 (0.05) | 0.09 (0.02) |
| CC/GG | -1.85 (0.02) | -1.82 (0.02) | **-1.84 (0.03)** | -2.08 | -1.93 (0.04) | 0.06 (0.02) |
| CG/GC | -2.27 (0.06) | -2.29 (0.10) | **-2.28 (0.12)** | -2.50 | -2.37 (0.09) | 0.13 (0.04) |
| GA/CT | -1.40 (0.07) | -1.63 (0.04) | -1.50 (0.08) | -1.57 | -1.47 (0.05) | 0.15 (0.02) |
| GC/CG | -2.30 (0.06) | -2.43 (0.10) | -2.36 (0.11) | -2.53 | -2.36 (0.04) | 0.08 (0.02) |
| TA/AT | -0.84 (0.08) | -0.87 (0.06) | -0.85 (0.10) | -0.84 | -0.84 (0.05) | 0.09 (0.02) |
| Loop | 2.30 (0.06) | 2.46 (0.09) | 2.37 (0.10) | 2.68 | 2.43 (0.05) | - |

**Table 1. Summary of results at 298 K, 1 M [NaCl]**

Free energies are given in kcal/mol. 6 kb (2 kb) are the energies obtained from the averaged results from the 6.8 kb (2.2 kb) sequences (standard error in parenthesis). Best is an average of the 2.2 kb and 6.8 kb results. In bold type letter we highlight the bps that disagree most with the values predicted by the UO model (extracted from ref. 2). $\varepsilon^o_i$ and $m_i$ are the standard energies and prefactors obtained from the fits of Eq. **3**, shown in Fig. 3.

| Method | UO Values | | | | Force measurements | | | |
|---|---|---|---|---|---|---|---|---|
| NNBP | $\varepsilon_i$ 25 ºC | $\Delta h_i$ | $\Delta s_i$ | $m_i$ | $\varepsilon_i$ 25 ºC | $\Delta h_i$ | $\Delta s_i$ | $m_i$ |
| AA/TT | -1.28 | -7.9 | -22.2 | 0.114 | -1.23 | **-7.28 (0.3)** | -20.28 (1.2) | 0.145 |
| AC/TG | -1.72 | -8.4 | -22.4 | 0.114 | -1.49 | **-5.80 (0.3)** | -14.46 (1.3) | 0.099 |
| AG/TC | -1.54 | -7.8 | -21.0 | 0.114 | -1.36 | **-5.21 (0.3)** | -12.89 (1.2) | 0.070 |
| AT/TA | -1.12 | -7.2 | -20.4 | 0.114 | -1.17 | **-4.63 (0.6)** | -11.62 (2.1) | 0.117 |
| CA/GT | -1.73 | -8.5 | -22.7 | 0.114 | -1.66 | **-8.96 (0.3)** | -24.48 (1.2) | 0.091 |
| CC/GG | -2.07 | -8.0 | -19.9 | 0.114 | -1.93 | **-8.57 (0.3)** | -22.30 (1.2) | 0.063 |
| CG/GC | -2.49 | -10.6 | -27.2 | 0.114 | -2.37 | **-9.66 (0.5)** | -24.43 (2.1) | 0.132 |
| GA/CT | -1.58 | -8.2 | -22.2 | 0.114 | -1.47 | **-8.16 (0.3)** | -22.46 (1.3) | 0.155 |
| GC/CG | -2.53 | -9.8 | -24.4 | 0.114 | -2.36 | **-10.10 (0.5)** | -25.96 (1.8) | 0.079 |
| TA/AT | -0.85 | -7.2 | -21.3 | 0.114 | -0.84 | **-8.31 (0.6)** | -25.06 (2.1) | 0.091 |
| $\chi^2$ | 2.37 | | | | 1.74 | | | |

**Table 2. Melting temperature prediction and optimal enthalpies and entropies**

Free energies and enthalpies given in kcal/mol; entropies given in cal/mol·K. UO values at standard conditions plus homogeneous salt corrections (left block of columns) have larger $\chi^2$ error in predicting melting temperatures for oligos longer than 15 bp of ref. 31 than heterogeneous salt corrections (right block of columns). Combining melting temperature data from ref. 31 and the new values for $\varepsilon^o_i$ and $m_i$ we can also extract the optimal enthalpies $\Delta h_i$ (highlighted) and entropies $\Delta s_i$. The values in parenthesis indicate the range of $\Delta h_i$ and $\Delta s_i$ that also predict the melting temperatures of the oligos within an average error of 2 ºC (i.e. the typical experimental error claimed in melting experiments).

Supporting Information of the paper

# Single-molecule derivation of salt dependent base-pair free energies in DNA

by

Josep M. Huguet[a], Cristiano V. Bizarro[a,b] , Núria Forns[a,b], Steven B. Smith[c], Carlos Bustamante[c,d], Felix Ritort[a,b,1]

[a]Departament de Física Fonamental, Universitat de Barcelona, Diagonal 647, 08028 Barcelona, Spain. [b]CIBER-BBN de Bioingenieria, Biomateriales y Nanomedicina, Instituto de Sanidad Carlos III, Madrid, Spain. [c]Department of Physics, [d]Department of Molecular and Cell Biology and Howard Hughes Medical Institute, University of California, Berkeley, CA 94720.

[1] To whom Correspondence should be addressed.
Facultat de Física, UB
Diagonal 647, 08028 Barcelona, Spain
Phone: +34-934035869
Fax: +34-934021149
E-mail: fritort@gmail.com, ritort@ffn.ub.es

# Contents

# S1. Optical Tweezers instrument

The experimental setup (see Figure S2) consists of two counter-propagating laser beams of 845 nm wavelength that form a single optical trap where particles can be trapped by gradient forces. The setup is similar to the one described by Smith *et al*. (1). Here two microscope objectives with numerical aperture 1.20 act as focusers and condensers simultaneously. One laser beam is focused through its objective while the other objective collects the exiting light, which is redirected to a position-sensitive detector (PSD). The laser beams have orthogonal polarizations thus making their optical paths separable by using polarized beamsplitters. When the optical trap exerts a force on a particle, the unbalanced exiting light is redirected from the back focal plane of the objective to a PSD that returns a current proportional to the force. Since the force is measured using conservation of light momentum, the calibration does not depend on the power of the lasers, the refraction index of the microspheres, their size or shape, the viscosity or refractive index of the buffer.

The optical trap can be positioned with the so-called wigglers. A wiggler is a device that bends the optical fiber of the laser using two piezoelectric crystals that mechanically push the fiber in such a way that the light is redirected to the desired position. There are two piezos per laser to place the optical trap at the desired location in the XY plane. The position of the center of the trap is measured by separating 5% of the light from the laser beam with a pellicle beam-splitter and forming a light-lever. For the lightlever we use a PSD in such a way that the measured current is proportional to the position of the beam. See refs. 2 and 3 for further information.

All the currents measured by the PSDs are processed by electronic microprocessors and the data are sent to a computer and converted to forces and distances. The acquisition frequency is 1 kHz and the resolution is 0.1 pN in force and 0.5 nm in distance. The experiments are carried out in a fluidics chamber that holds the micropipette. The whole chamber can be moved with a motorized stage. The instrument and the surrounding room are kept at a constant temperature of 25 ± 0.3 ºC.

## S2. Force and distance calibration

According to the experimental setup (Sec. S1), the instrument detects the change in the light momentum of the laser beams that form the optical trap, which allows us to directly measure the force accurately. There is a linear relation between the PSD reading and the actual force exerted on the bead by the optical trap:

$$F_y = M \cdot PSD_y + F_o \qquad \textbf{[1]}$$

where $F_y$ is the actual force on the *y*-axis in pN, $PSD_y$ is the sum of the readings of the PSDs of both traps in the *y* direction in adu (analog to digital units), *M* is the calibration factor and $F_o$ is an offset, already corrected by the data acquisition board. The process of calibration consists on accurately find out the value of *M*, which is independent of the trap power. Here we only show the method used in force calibration for the *y*-axis. The same procedure applies to *x* and *z* axis. Three different methods were used to calibrate the PSD that measures the force.

In the first method, we collected 30 s of Brownian $PSD_y$ signal for a trapped microsphere in distilled water at low laser power (Figure S3a), in order to have a low corner frequency. The signal was squared and a power spectrum built up (Figure S3b) by averaging the spectra of 1.00-second windowed frames of the data. The averaged spectrum was fit in log-log scale (Figure S3c) to a Lorentzian profile:

$$S_{PSD_y}(\nu) = \left\langle PSD_y(\nu) \cdot PSD_y^*(\nu) \right\rangle = \frac{A}{B + (2\pi\nu)^2} \qquad \textbf{[2]}$$

where $S_{PSD_y}(\nu)$ is the power density of the $PSD_y$ noise in adu units and A and B are the fit parameters. The power density of the force noise of a trapped particle (1) is expected to follow a Lorentzian distribution according to:

$$S_{F_y}(\nu) = \left\langle F_y(\nu) \cdot F_y^*(\nu) \right\rangle = \frac{2k\, k_B T\, \omega_c}{\omega_c^2 + (2\pi\nu)^2} \quad ; \quad \omega_\mathrm{c} = \frac{k}{\gamma} \qquad \textbf{[3]}$$

where $S_{F_y}(\nu)$ is the force power density of the *y*-axis force, *k* is the trap stiffness in the *y* direction, $k_B$ is the Boltzmann constant, *T* is the temperature, $\omega_c$ is the corner frequency and $\gamma$ is the drag coefficient of the bead in distilled water. The drag coefficient for spherical particles at low Reynolds number regime can be calculated according to $\gamma=6\pi\eta R$, where $\eta$ is the viscosity of distilled water at 25ºC and *R* is the bead radius. The viscosity of water at 25ºC is taken as $\eta=8.9\cdot10^{-4}$ Pa·s and the diameter of the bead is taken as *R*=3.00±0.05 μm from scanning electron microscopy measurements (average over 100 3.0-3.4 μm polystyrene beads from Spherotech, Libertyville, IL). Using Eq. **1**, the measured power spectrum (in adu units) and the expected one (in pN) can be related:

$$\begin{aligned} S_{F_y}(\nu) &= M^2 \cdot S_{PSD_y}(\nu) \\ \frac{2k\, k_B T\, \omega_c}{\omega_c^2 + (2\pi\nu)^2} &= \frac{M^2 A}{B + (2\pi\nu)^2} \end{aligned} \qquad \textbf{[4]}$$

from which the calibration factor (*M*) and the trap stiffness (*k*) can be obtained by identifying the parameters of the fit spectrum with the expected one:

$$\left.\begin{aligned} 2k\, k_B T\, \omega_c &= M^2 A \\ \omega_c^2 &= B \end{aligned}\right\} \qquad \left.\begin{aligned} M &= \sqrt{2\, k_B T\, \gamma B / A} \\ k &= \gamma\sqrt{B} \end{aligned}\right\} \qquad \textbf{[5]}$$

The two force calibration factors extracted from the power spectra measured at two different laser powers differed less than 1%. The stiffness of the weak trap was *k*=3.22±0.05 pN/μm, while *k*=6.20±0.07 pN/μm for the stronger trap. A first check of the correctness of these numbers was made by calculating the trap stiffness from the force fluctuations (i.e. the PSD reading) of previous measured time series (Figure S3a). The fluctuation-dissipation relation for a trapped particle can be written as:

$$\left\langle F_y^2 \right\rangle = k \cdot k_B T \qquad \textbf{[6]}$$

where $\left\langle F_y^2 \right\rangle$ is the variance of the *y* force. Since we have already calculated the calibration factor (*M*) we can combine Eq. **6** and **1** to write:

$$k = \frac{M^2 \left\langle PSD_y^2 \right\rangle}{k_B T} \qquad \textbf{[7]}$$

where $\left\langle PSD_y^2 \right\rangle$ is the variance of the PSD in the *y* direction. In this way we can calculate the trap stiffness without knowing the viscosity of water. The values obtained were $k = 3.26$ pN/nm for the soft trap and $k = 6.15$ pN/nm for the stiff trap, which represents an error of 1.5%. Once the optical trap was calibrated, a check was performed to measure the stiffness using another independent method (Figure S3d) which does not use the force noise. A bead was stuck at the tip of the micropipette. The optical trap was aligned in the center of the bead and short displacements of the optical trap were performed while recording the force vs. trap displacement and keeping the micropipette at fixed position. The slope of the linear region of this curve is the trap stiffness. The values of the stiffness obtained were $k = 3.200\pm0.002$ pN/μm for the soft trap and $k = 6.28\pm0.02$ pN/μm for the stiff trap. These values differ by less than 1.5% with respect to the power spectrum method.

In the second method, the PSDs were calibrated using Stokes law (Figure S3e). A microsphere was trapped and the whole fluidics chamber moved at a fixed speed using the motorized stage. The microsphere undergoes a force produced by the surrounding fluid which is proportional to the speed. Knowing the microsphere radius and the viscosity of distilled water, the relation between the PSDs readings and the velocity of the fluid gives the calibration factor. The calibration factor is obtained from averaging 40 different beads (same beads as above). The calibration factor obtained agreed within 2% with the previous methods.

A third method that does not depend on the bead size and viscosity of water was used to check our previous calibration protocols. A similar setup to Figure S2 (instrument in Barcelona) was also made in Berkeley and calibrated using the known momentum properties of light. It can be shown (1) that the force sensitivity of the PSD detector is given by

$$M = (R_D / f_O)/(\Psi \cdot c) \quad \textbf{[8]}$$

where $R_D$ is the half-width of the PSD chip, $f_o$ is the objective lens focal length, $\Psi$ is the power sensitivity of the PSD (signal/watts referenced to the trap position) and $c$ is the speed of light. The PSD responsivity was measured with an optical power meter (Thorlabs PM30-130) and PSD dimensions were tested by using a test laser on a motorized stage (4). Using this calibration, we pulled the same 2.2 kb molecule in Berkeley and obtained the data shown in Figure S4 where we see good agreement between the FEC of the 2.2 kb molecule measured in different instruments. The mean unzipping force differs by less than 0.15 pN.

The calibration of the lightlever position sensor is done using the motors that move the XYZ-stage. The Thorlabs Z-606 motors have a shaft encoder that counts the turns of the axes so that the position of the stage can be determined. A trapped microsphere is held fixed at the tip of the micropipette and the optical trap follows the position of the microsphere by keeping the total force equal to zero using a force feedback mechanism (Figure S5a). When the micropipette moves, the wiggler acts to reposition the center of the trap in order to follow the bead. The calibration factor for the position can be determined through the relation between the reading of the position PSDs (also known as lightlevers) and the position of the motor. This calibration protocol provides slightly different calibration factors (less than 3% difference) depending on the direction in which the motor is moving, due to the backslash of the motor (Figure S5b,c).

Summing up, we verified that our instrument is well calibrated in force within 0.5 pN (i.e. ~3%) and in distance within 3%.

## S3. DNA molecular constructs

A 6770 bp insert DNA was isolated by gel extraction of a *BamH*I digestion of λ phage DNA (see Figure S6). Two short handles of 29 bps and one tetraloop (5'-ACTA-3') were ligated to the insert that has the λ cosR end and a *BamH*I sticky end. To construct the DNA handles, an oligonucleotide (previously modified at its 3' end with several digoxigenins using DIG Oligonucleotide Tailing Kit, 2$^{nd}$ Generation, Roche Applied Science) was hybridized with a second 5' biotin-modified oligonucleotide giving a DNA construction with one cohesive end complementary to cosR and two 29 nucleotide long ssDNA at the other end. These two ssDNA have the same sequence and they were hybridized with a third oligonucleotide, which is complementary to them, resulting in two dsDNA handles. This construction was attached to the insert DNA by a ligation reaction. A fourth self-complementary oligonucleotide, which forms a loop in one extreme and a cohesive *BamH*I end at the other, was ligated to the *BamH*I sticky end of the insert DNA. The DNA was kept in aqueous buffer containing 10 mM Tris-HCl (pH 7.5) and 1 mM EDTA. Streptavidin-coated polystyrene microspheres (2.0-2.9 μm; G. Kisker GbR, Products for Biotechnologie) and protein G microspheres (3.0-3.4 μm; Spherotech, Libertyville, IL) coated with anti-digoxigenin polyclonal antibodies (Roche Applied Science) were used for specific attachments to the DNA molecular construction described above. Attachment to the anti-digoxigenin microspheres was achieved first by incubating the beads with tether DNA. The second attachment was achieved in the fluidics chamber and was accomplished by bringing a trapped anti-digoxigenin and a streptavidin microsphere close to each other.

The 2.2 kb construct was obtained taking the 2215 bp fragment from a *Sph*I digestion of λ DNA. The same two short handles and tetraloop used for the 6.8 kb DNA are used for the 2.2 kb construct, with the following exceptions: the two handles are hybridized to the 2215 bp DNA through the cosL cohesive end of λ DNA, and the tetraloop was added to the insert DNA using the *Sph*I sticky end. Table S3 shows the fraction of the 10 NNBPs found along the two sequences.

## S4. Synthesis of ssDNA

A 3 kb ssDNA molecular construct was obtained by pH denaturation (strand separation) of a 3 kb dsDNA (see Figure S7). The dsDNA was obtained from PCR amplification of a ~3 kb fragment of λ-DNA. One of the primers used in the process was already labeled with Biotin. The resulting product was cleaved with the endonuclease *Xba*I producing a cohesive end. Another 24 base oligonucleotide (previously labeled with several digoxigenins at its 3' end by using terminal transferase) was hybridized with a second 20-base long oligonucleotide giving a DNA construction with one cohesive end complementary to *Xba*I. Both products were annealed and ligated resulting in one 3 kb dsDNA molecule. To produce ssDNA, the molecular construct was incubated with Streptavidin coated beads for 30 min at room temperature in a volume of 15 μl of 10 mM NaCl TE buffer. Afterwards, 35 μl of 0.1 M NaOH were added in order to cause the separation (i.e. denaturation) of the strands. After 30 min, the sample was centrifuged. The white precipitate of beads and ssDNA was re-suspended in TE buffer. The second attachment with the antidigoxigenin beads was achieved in the fluidics chamber with the help of the micropipette.

## S5. Elastic parameters of ssDNA and comparison between the two models

From the pulling experiments on ssDNA (see Fig. 2e in main text and Figure S8) we conclude that the Worm-Like Chain (WLC) model correctly describes the elastic response of the ssDNA at low salt concentration (<100 mM [NaCl]) whereas the Freely-Jointed Chain (FJC) model works better at higher salt concentrations (>100 mM [NaCl]). Strictly speaking, in the later regime both models fail because the elastic response of the ssDNA exhibits a force plateau at low forces (Fig. 1e in main text or rightmost panel in Figure S8). Such force plateau indicates that the ssDNA has some structure at high salt concentration and low forces. The shape of the force plateau is hardly reproducible and varies from pulling to pulling and from molecule to molecule. Therefore, it is quite difficult to establish which is the native state of a ssDNA molecule at low forces. Nevertheless, our goal is to measure the free energy difference between two complementary ideal ssDNA chains and the duplex of dsDNA that they form. So we do not need to know the elastic response of a self-interacting ssDNA molecule. Instead, we assume that the ssDNA behaves like an ideal chain.

In order to obtain the elastic response of the ssDNA, we have to fit the stretching experimental data of the 3 kbp ssDNA molecules to the appropriate model (FJC or WLC depending on the salt concentration). At high salt concentration, any pulling FDC of a ssDNA molecule (like in Fig. 1e main text) will be correctly fit by a FJC model above 15 pN because no secondary structure will survive at this force. Therefore, we can fit the FDCs above 15 pN to a FJC model. Note that we are clearly neglecting any possible secondary structure that can be formed below 15 pN, because the FJC model itself is an ideal chain and it does not account for such structures. Figure S8 shows the elastic response of the ssDNA for three different salt conditions together with the best fits to the two elastic models (WLC and FJC). Table S4 summarizes our results for the elastic parameters.

The fit is performed as follows. The measured FDCs of the ssDNA (right panels Fig. 2e main text and Figure S8) are converted to a force-extension curve (FEC) after subtracting the contribution of the optical trap according to,

$$x_s = x_{\mathrm{tot}} - \frac{f}{k}$$

where $x_s$ is the extension of the ssDNA, $x_{\mathrm{tot}}$ is the total distance, $f$ is the force and $k$ is the stiffness of the optical trap. The resulting FEC is fit to a FJC or a WLC (see upper panels in Figure S9). The fitting parameters are the Kuhn length ($b$) or the persistence length ($l_p$) depending on the model, and the interphosphate distance ($d$). The FEC is forced to pass through the point ($x_s$=0, $f$=0), while the number of bases is fixed to $n$=3000 (see "Synthesis of ssDNA" in Materials and Methods section in main text).

As a final test, the fit Kuhn length (or persistence length) of the ssDNA is introduced into the equation of the total distance of the system (see "Calculation of the equilibrium FDC" in the Materials and Methods section in main text). When the molecular construct is fully unzipped, it is actually a ssDNA molecule tethered between two beads (plus two short dsDNA handles). So the last part of the unzipping FDC (above ≅15 pN) is giving us the elastic response of the ssDNA. Here, the ssDNA shows its elastic response because no secondary structure can be formed. Lower pannels in Figure S9 show that the last part of the unzipping FDCs are correctly described by a ssDNA molecule of 2×6838 bases.

In the end, we have fit the elastic response of the ssDNA to a FJC above 15 pN. Below this force, we have assumed that the FJC correctly describes the ideal elastic response of the ssDNA, i.e. in the absence of secondary structure.

## S6. Monte Carlo minimization algorithm

We developed an algorithm based on Monte Carlo (MC) minimization where we start from an initial guess for the energies $\varepsilon_i$ ($i$=1,..,10) and do a random walk in the space of parameters in order to minimize the error function (Eq. **2** main text). The method is just a standard Simulated Annealing optimization algorithm (5) adapted to our particular problem that speeds up considerably the time to find the minimum as compared to standard Steepest Descent algorithms. The error landscape defined by Eq. **2** main text is not rough and there is no necessity to use a MC algorithm. Nevertheless, we find that the MC optimization is computationally more efficient (i.e. faster) than other optimization algorithms (such as Steepest Descent) because no derivatives need to be calculated. A fictive temperature is defined to control how the space of parameters is explored. Each new proposal is accepted or rejected using a Metropolis algorithm. According to the Metropolis algorithm a move is always accepted if $\Delta E < 0$ (where $\Delta E$ is the change in the error after the proposal). If $\Delta E > 0$, a random number $r$ from a standard uniform distribution $r \in U(0,1)$ is generated and the move accepted if $e^{-\Delta E/T} > r$ and rejected if $e^{-\Delta E/T} < r$. A *quenching* MC protocol, where the minimization algorithm is run at a very low fictive temperature, was found to be particularly efficient.

We start from the unified values for the NNBP energies to enforce the system to explore the basin of attraction in the vicinity of the unified values. The system is allowed to evolve until the total error reaches a minimum. This method is essentially a steepest descent algorithm with the advantage that the free energy derivatives need not be calculated (see Figure S10a). Once the system has found the minimum, we start another MC search to explore other solutions in the vicinity of the minimum. For that we use a *heat-quench* algorithm (see Figure S10b) in which the system is heated up to a large fictive temperature until the error is 50% higher than the error of the first minimum. Afterwards, the system is quenched until the acceptance of MC steps is lower than 0.03%. This procedure is repeated many times until the multiple solutions allow us to estimate the error of the algorithm. The possible values for the stacking energies are Gaussian distributed in a region of width approximately equal to 0.05 kcal/mol calculated (see Figure S10c). The analysis revealed that the optimization algorithm is robust and leads to the same solution when the initial conditions are modified (see Sec. S13). Different molecules and different sequences converge to energy values that are clustered around the same value (see Sec. S14).

# S7. Drift and shift function

Instrument drift is a major problem in single molecule experiments. The drift is a low frequency systematic deviation of measurements due to macroscopic effects. The importance of drift depends on the kind of the experiment and the protocol used. Our unzipping experiments are performed at very low pulling speeds (typically around 10 nm/s), so measuring a whole unzipping/rezipping FDC may take 10 minutes or longer. Therefore it is useful to model drift in order to remove its effects and extract accurate estimates for the NNBP energies.

Our experimental measurements are force versus trap position as measured by the lightlevers (Sec. S1). The unzipping/zipping curves contain reproducible and recognizable landmarks (i.e. slopes and rips) which indicate the true position of the trap. Therefore we devised a way to take advantage of these landmarks to correct for the instrumental drift.

Correction for drift is introduced in terms of a *shift* function $s(x_{tot})$ which is built in several steps. Due to its relevance for data analysis we describe the steps in some detail:

Step 1. We start with the experimental FDC filtered at 1 Hz bandwidth and fix the origin of coordinates for the distance $x_{tot}$ (trap distances are relative) by fitting the last part of the experimental FDC (black and magenta curves in Figure S11a) corresponding to the stretching of the ssDNA when the hairpin is fully unzipped.

Step 2. Having fixed the origin of coordinates we calculate the predicted FDC by using the Unified Oligonucleotide (UO) NNBP energies. It is shown in red (Figure S11a, S11b). The qualitative behavior is acceptable (all force rips are reproduced). However, the predicted mean unzipping force is higher than the value found experimentally and the force rips are not located at the correct position.

Step 3. Next we generate a FDC with NNBP energies lower than the UO energies until the mean unzipping forces of the predicted and the experimental FDC coincide. Typically what we do is multiplying all the 10 NNBP UO energies by a factor ~0.95. The new NNBP energies have an absolute value 8-10% lower than the UO NNBP energies. The resulting FDC with these new energies is shown in green in Figure S11b (in this particular case we took $\varepsilon_i^{\mathrm{New}} = 0.92 \cdot \varepsilon_i^{\mathrm{Mfold}}$). Although the mean unzipping force of the green and black curves is nearly the same, there is misalignment between the rips along the distance axis. Moreover, there are discrepancies between the predicted and measured heights of the force rips. As we will see below, the shift function will correct the horizontal misalignments and the NNBP energies will correct the discrepancies along the force axis.

Step 4. We now introduce a shift function that uses the slopes preceding the rips as landmark points to locally correct the distance to align the experimental data with the theoretical prediction. First, we want to know the approximate shape of the shift function and latter we will refine it. This is done by looking for some characteristic slopes of the sawtooth pattern along the FDC and measuring the local shift that would make the two slopes (theoretical and experimental) superimpose. Figure S11c shows zoomed regions of the FDC and the blue arrows indicate the local shift that should be introduced in each slope to correct the FDC. The orange dots shown in the upper panel of Figure S11d depict the local shifts vs. the relative distance that have been obtained for the landmark points. These orange dots represent a discrete sampled version of an ideal shift function that would superimpose the predicted and the experimental FDC. Because these dots are not equidistant, we use cubic splines to interpolate a continuous curve every three landmark points. The resulting interpolated function that describes the local shift for any relative distance is shown in violet. Note that the violet curve passes through all the orange dots.

Step 5. Starting from the cubic splines interpolation of the shift function that we have found (violet curve in upper panel Figure S11d) we can define new equidistant points (yellow dots in lower panel Figure S11d) that define the same shift function. The yellow equidistant points are separated 100 nm. We call these yellow points “Control points”.

Step 6. We now introduce the shift function into the calculation of the theoretical FDC. The results are shown in Figure S11e. Again, the black curve is the experimental FDC, the green curve is the predicted FDC without the local shift correction and the magenta curve is the predicted FDC with the local shift function obtained previously. Note that the magenta and green curves are identical, except for local contractions and dilatations of the magenta curve. The slopes of the magenta and the black curves now coincide. Still, the NNBP energies must be fit to make the height of the rips between the theoretical and experimental curve coincident.

Step 7. At this point we start the Monte Carlo fitting algorithm. At each Monte Carlo step we propose new values for the 10 NNBP energies and we also adjust the shift function in order to superimpose the theoretical and experimentally measured FDCs. The shift function is adjusted by modifying the values of the control points (yellow dots in lower panel of Figure S11d). The horizontal position of the yellow points is always the same (e.g. the yellow dot located at Relative Distance = -1000 nm will always be located there). What we change when we adjust the shift function is the value of each point (e.g. the yellow dot located at Relative Distance = -1000 nm may change its shift value from -54 to -20 nm).

During the Monte Carlo optimizing procedure the shift function modifies its shape as the NNBP energies are modified. The results are shown in Figure S11f. Black curves show snapshots of the evolution of the shift function during the optimization process from the initial shape (red curve). The green curve shows the final shift function (the control points are not depicted in Figure S11f, only the interpolated shift function). When we finally calculate the theoretical FDC using the optimal shift function and the optimal NNBP energies we get the maximum overlap between the theoretical prediction and the experimental FDC. Black curve in upper panel in Figure S11g is the experimental FDC and red curve is the predicted FDC after having fit the NNBP energies and the shift function. The correction for drift has now finished. The optimal shift function is also shown in the lower panel in Figure S11g.

All the steps we described before were applied to all different molecules and salt conditions we measured. Measurements from different molecules have shift functions of similar shapes (Figure S12). We have checked that the undulations are not an artifact of the spline interpolation. The undulations remain when the number of control points of the shift function is increased. The net shift observed in some curves (around ±100 nm) might be explained by improper calibration of the distance (around 4%). The undulations observed in the shift function might be due to non-linearities in the lightlever (i.e. trap position) measurements or interference fringes in the lenses and the pellicle located along the optical path to the PSDs. The undulations in the shift function might also be correlated with the DNA sequence as emerges from the fact that undulations observed in different molecules of the same sequence appear at nearby positions. This might indicate new effects in the unzipping curves not accounted for in the NN model (e.g. the presence of next nearest neighbor corrections).

We have also checked whether the correction introduced by the shift function also could be explained by a dependence of the Kuhn length on the contour length. To our knowledge, such dependence has not ever been reported. Yet it is interesting to evaluate the consequences of such hypothetic dependence. By letting the Kuhn length depend on the number of open base pairs, the position of the theoretical and experimental slopes and rips match each other if the Kuhn length increases as the contour length decreases. However this matching occurs at the price of an increasing average mean unzipping force as the molecule unzips and the ssDNA is released, an effect which is not experimentally observed. We conclude that the shift function is probably due to instrumental drift superimposed to imperfect calibration of the distance and non-linear optical effects.

## S8. Calculation of melting temperatures

The melting temperature of a DNA hairpin is defined as the temperature at which half of the molecules are in the native state (i.e. the double helix state) and half are denaturated (i.e. the two strands are split). The nearest-neighbor model is widely used to predict melting temperatures of DNA duplexes. In order to predict melting temperatures the enthalpy and entropy of formation of the DNA hairpin must be known. The melting temperature of a non-self-complementary duplex is given by (6):

$$T_M = \frac{\Delta H^o}{\Delta S^o + R\ln[C_T/4]} \qquad \textbf{[9]}$$

where $\Delta H^o$ and $\Delta S^o$ are the enthalpy and entropy of formation at 1 M NaCl respectively and they are assumed to be independent of the temperature, $R$ is the ideal gas constant (1.987 cal/K·mol), $[C_T]$ is the total oligonucleotide strand concentration of DNA molecules and the factor ¼ must be included for non-self-complementary molecules. For each oligo, $\Delta H^o$ (and $\Delta S^o$) has two contributions: 1) the NNBP contribution and 2) the initiation term. The initiation term depends on the first and the last base pair of the oligo. We take the values of the initiation terms from ref. 6 and we assume that they do not depend on the temperature nor the salt concentration (see Table S1). To extend the prediction of the melting temperatures to different salt conditions, the salt dependence of the entropy has to be considered. The Unified Oligonucleotide (UO) model predicts the melting temperature at different salt conditions assuming a homogeneous correction for all the 10 NNBP. At a salt condition $[Mon^+]$ (where $[Mon^+]$ is the total concentration of monovalent ions) different from 1 M NaCl, the UO model corrects $\Delta S^o$ according to

$$\Delta S^0([\mathrm{Mon}^+]) = \Delta S^0(1\ \mathrm{M\ NaCl}) + 0.368 \times N \times \ln[\mathrm{Mon}^+] \qquad \textbf{[10]}$$

where $N$ is the number of phosphates divided by 2 (i.e. the number of base pairs of the hairpin). Our heterogeneous salt correction assumes a different prefactor for each NNBP, which is temperature independent. Therefore, the entropy of an oligo is corrected according to

$$\Delta S^0([\mathrm{Na}^+]) = \Delta S^0(1\ \mathrm{M\ NaCl}) + \sum_{i=1}^{N} \frac{m_i(T)}{T} \ln[\mathrm{Mon}^+] \qquad \textbf{[11]}$$

where $m_i(T)$ are the specific salt corrections at $T$ = 298 K and they have to be summed over all $N$ base pairs of the hairpin. Note that the prefactor $m_i(T)/T$ is independent of the temperature. Following this scheme we find that our heterogeneous salt correction gives the following prediction for the melting temperatures,

$$T_M = \frac{\Delta H^0}{\Delta S^0 + \sum_{i=1}^{N} \frac{m_i(T)}{T} \ln[\mathrm{Mon}^+] + R \ln\left[C_T/4\right]} \qquad \textbf{[12]}$$

which is Equation 4 of the main text.

# S9. Enthalpy and entropy inference

Our unzipping experiments provide direct measurements of the free energies ($\varepsilon_i$) and the salt correction ($m_i$) at $T$ = 298 K for all the NNBP ($i$ = 1,...,10), but no information about the enthalpy ($\Delta h_i$) and the entropy ($\Delta s_i$) is provided. However, combining our results with the measurements of melting temperatures of several oligos obtained by optical melting experiments we can infer the enthalpies and the entropies. In order to do so, we define an error function ($\chi^2$) that accounts for the mean squared error between the experimental melting temperatures (7) ($T_i^{exp}$) and the predicted ($T_i^{pred}$) ones for $N$ different oligos and salt conditions,

$$\chi^2(\Delta h_1,\ldots,\Delta h_{10}) = \tfrac{1}{N}\sum_i^N \left(T_i^{exp} - T_i^{pred}(\Delta h_1,\ldots,\Delta h_{10})\right)^2 \qquad \textbf{[13]}$$

where $\Delta h_i$ ($i$=1,...,10) are the NNBP enthalpies and $T_i^{pred}$ are obtained according to Eq. **12** The NNBP entropies are fixed by 10 constraints that relate the free energies, the enthalpies and the entropies according to

$$\varepsilon_i = \Delta h_i - T\Delta s_i \quad \Rightarrow \quad \Delta s_i = \frac{\Delta h_i - \varepsilon_i}{T} \qquad \textbf{[14]}$$

where $i$ = 1,...,10; $\varepsilon_i$ are the experimentally measured free energies with unzipping and $T$ = 298 K. Here, the enthalpies are fitting parameters that fix the entropies. Therefore the enthalpies and the entropies are fully correlated (their correlation coefficients are equal to 1). The error function is minimized with respect to the enthalpies using a steepest descent algorithm that rapidly converges to the same solution when starting from different initial conditions.

Here we provide and estimation of the error ($\sigma_{\Delta h_i}$) of the 10 fitting parameters $\Delta h_i^0$ ($i$ = 1,...,10). We simplify the notation by writing the $\Delta h_i^0$ ($i$=1,...10) values that we give in Table 2 main text in vectorial form according to $\Delta\vec{h}_m$, where $m$ stands for minimum. Note that $\Delta\vec{h}_m$ minimizes the $\chi^2(\Delta\vec{h})$ error function (Eq. **13**). By definition, the first derivatives of $\chi^2(\Delta\vec{h})$ with respect to $\Delta\vec{h}$ vanish at the minimum ($\vec{\nabla}\cdot\chi^2(\Delta\vec{h}_m)=0$). So we can write a Taylor expansion of $\chi^2(\Delta\vec{h})$ up to second order according to:

$$\chi^2(\Delta\vec{h}_m + \delta\Delta\vec{h}) \approx \chi^2(\Delta\vec{h}_m) + \frac{1}{2}\delta\Delta\vec{h}^T \cdot H\left(\chi^2(\Delta\vec{h}_m)\right)\cdot\delta\Delta\vec{h} \qquad \textbf{[15]}$$

where $\delta\vec{\Delta h}$ is a variation of the $\vec{\Delta h_m}$ vector and $H(\chi^2(\vec{\Delta h_m}))$ is the Hessian matrix of second derivaties $H_{ij} = \dfrac{\partial^2 \chi^2(\Delta h_1^0,...,\Delta h_{10}^0)}{\partial \Delta h_i^0 \, \partial \Delta h_j^0}$ evaluated at the minimum $\vec{\Delta h_m}$. Our estimation of $\chi^2(\vec{\Delta h_m}) = 1.74$ (in units of squared Celsius degrees ºC$^2$) is lower than the typical experimental error in melting experiments, which is 2 ºC (i.e $\chi^2 = 4$). So there is a range of $\vec{\Delta h}$ values around the minimum $\vec{\Delta h_m}$ that still predict the melting energies within an average error of 2 ºC. This range of values is what determines the error in the estimation of $\vec{\Delta h_m}$. Following this criterion, we look for the variations around the minimum ($\delta\vec{\Delta h}$) that produce a quadratic error of 4 ºC$^2$. We divide this quadratic error into the 10 fitting parameters ($\Delta h_i^0$ $i$=1,…,10) and the 10 related ones ($\Delta s_i^0$ $i$=1,…,10). So we look for each $\delta\Delta h_i$ that induces an error or 4 ºC$^2$/20 = 0.2 ºC$^2$. Now, introducing $\chi^2(\vec{\Delta h_m} + \delta\vec{\Delta h}) = 4$ and $\chi^2(\vec{\Delta h_m}) = 1.74$ into Eq. **15** and isolating $\delta\vec{\Delta h}$, we get one expression to estimate the errors of the 10 fitting parameters:

$$\delta\Delta h_i = \sigma_{\Delta h_i} = \sqrt{\frac{2\cdot(0.2-0.087)}{H_{ii}(\chi^2(\vec{\Delta h_m}))}} = \sqrt{\frac{0.226}{H_{ii}(\chi^2(\vec{\Delta h_m}))}}, \qquad i = 1,...,10 \qquad \textbf{[16]}$$

which gives values between 0.3 - 0.6 kcal/mol (see Table 2 main text). Now, the error in the estimation of the entropies ($\sigma_{\Delta s_i}$) can be obtained from error propagation of Eq. **14**:

$$\sigma_{\Delta s_i} = \frac{1}{T}\left(\sigma_{\Delta s_i} + \sigma_{\Delta \varepsilon_i}\right) \qquad i = 1,...,10 \qquad \textbf{[17]}$$

where $T$ = 298.15 K is the temperature and $\sigma_{\Delta \varepsilon_i}$ are the experimental errors of our estimated NNBP energies from the unzipping measurements (Table 1, main text). The errors range between 1.2 - 2.2 cal/mol·K (Table 2, main text). Our results are compatible with the UO enthalpies and entropies.

## S10. Free energy of the loop

The end loop of the molecule is a group of 4 bases that forms a structure that facilitates the rezipping of the two strands of the dsDNA hairpin. The free energy formation of the loop gets contributions from the bending energy, the stacking of the bases in the loop and the loss of entropy of the ssDNA. The energy formation of the loop is positive, meaning that the loop is an unstable structure at zero force. Upon decreasing the total extension, the formation of complementary base pairs along the sequence reduces the total energy of the molecule and the loop can be formed.

The effect of the loop is appreciated only in the last rip of the FDC. It introduces a correction to the free energy of the fully extended ssDNA molecule and modifies the force at which the last rip is observed (Figure S13).

## S11. Sampling of energy states distribution

According to the NN model, the energy of the DNA duplex is higher when more base pairs are open. In our experiments, the opening of base pairs is sequential, meaning that the base pair that is closest to the opening fork is the one that dissociates first. As the molecule is pulled, the unzipping fork that separates the ssDNA from the dsDNA progressively advances as more dsDNA is converted into released ssDNA (see Figure S14a). This can be achieved thanks to the stiffness of the parabolic trap potential, which is high enough to allow us to progressively unzip the molecular construct and access to any particular region of the sequence. The position of the unzipping fork is determined by the number of open base pairs, which minimizes the total energy of the system at a fixed distance (i.e. at a fixed trap position). In general, the position of the unzipping fork (even at a fixed distance) exhibits thermally induced fluctuations in such a way that the system can explore higher free energy states. Such fluctuations represent the first kind of excitations in the system and will be discussed in the next paragraphs. However there is a second kind of excitation: breathing fluctuations. The breathing is the spontaneous opening and closing of base pairs produced in the dsDNA, far away from the unzipping fork (see Figure S14b). During this process, the DNA explores states of higher free energy while the unzipping fork is kept at the same position. So the breathing does not induce any change in the position of the unzipping fork. Consequently, we are not able to distinguish the breathing in our unzipping experiments because breathing fluctuations are not coupled to the reaction coordinate that we measure, i.e. the molecular extension, and should have a small effect on the measured FDC. Note that breathing fluctuations are expected to be relevant only at high enough temperatures. While the inclusion of breathing fluctuations should be considered at high enough temperatures their contribution at 25ºC is expected to be minimal. The fact that our model reproduces very well the experimental FDC supports this conclusion.

Now let us focus on the fluctuations of the unzipping fork. As explained in the manuscript, the unzipping of DNA is performed at very low pulling rate in our experiments. The pulling process is so slow that the system reaches the equilibrium at every fixed distance along the pulling protocol. Figure S15a shows a fragment of the FDC in a region where 3 states having different number of open base pairs coexist ($n_1$=1193, $n_2$=1248 and $n_3$=1300). Figure S15b shows the hopping in force due to the transitions that occur between these 3 states. The slow pulling rate guarantees that the hopping transitions are measured during unzipping (i.e. many hopping events take place while the molecule is slowly unzipped). The filtering of the raw FDC data produces a reasonably good estimation of the equilibrium FDC. It is also important to remark that the unzipping and rezipping curves are reversible (see Fig. 1c main text). This supports the idea that the unzipping process is quasistatic and correctly samples the energy states.

In general, the hopping frequency between coexistent states is around ~10 - 50 Hz and the area of coexistence extends over 40 nm of distance. At a pulling rate of 10 nm/s, we can measure around 10-40 transitions, which in most cases is sufficient to obtain a good estimation of the FDC after averaging out the raw data.

Figure S16a shows the free energy landscape of one molecule at different fixed distances, which is given by $G(x_{\mathrm{tot}},n)$ in Eq. 1 main text (see also the subsection *Calculation of the* *equilibrium FDC* in the *Materials and Methods* section in main). A detailed view of the free energy landscape (see Figure S16b) shows that it is a rough function and its coarse grained shape is parabolic. It means that for each value of $x_{\mathrm{tot}}$ there is always a state of minimum global energy surrounded by other states of higher free energies (see Figure S16b). Although there are lots of states in phase space, in the experiments we only observe those states that differ in free energy by less than ~5 $k_BT$ with respect to the state of minimum free energy. So the hopping transitions described in Figure S15 are between states that have similar free energies. Outside this range of free energies, the higher energetic states are rarely observed and their contribution to the equilibrium FDC is negligible.

Summing up, the sufficiently high trap stiffness, the short length of the molecular handles, the slow pulling rate and the shape of the free energy landscape ensure us that we explore higher energetic states (within a range of ~5 $k_BT$ with respect to the global minimum) during the unzipping process. Therefore, the averaged FDC is a good estimation of the equilibrium FDC.

## S12. Thermodynamics of DNA unzipping

The state of the system is determined by the number of open bp and the position of the center of the trap that fixes the total distance of the system. These two parameters and the force allow us to understand the process the molecule undergoes in an unzipping experiment. A useful three-dimensional representation of the space of variables is shown in Figure S17. The experiment starts when the molecule is closed and relaxed. We call this state A, in which the distance, the force and the number of open bps are equal to 0. The molecule unzips as the total distance increases (red curve). When the molecule is fully extended and stretched, the system is in the state B. The state C corresponds to a relaxed random coil of ssDNA. The free energy of formation of the duplex is equal to the sum of all NNBP energy contributions along the sequence and is given by the difference in free energy between states A and C. C is an inaccessible experimental state because ssDNA forms secondary structures at low forces. However, C can be recovered theoretically from B by considering that the fully extended molecule behaves like a random coil without interactions and is accurately described by an elastic polymer model.

## S13. Dependence of the optimization algorithm on the initial conditions

We have carried out a detailed study of the optimization algorithm in order to check that the final solution does not depend on the initial conditions given to the algorithm. An ensemble of initial conditions where selected from the values of the different labs that were unified by SantaLucia (6) (Figure S18a). The FDCs predicted by the different energy values are depicted in Figure S18c. Here we can observe how an overestimation (underestimation) in absolute value of the NNBP energies leads to an overestimation (underestimation) of the mean unzipping force. The same elastic properties (ssDNA, handles and optical trap) have been used in all cases. Figure S18b shows the optimal values of the NNBP energies for the ensemble of initial conditions. All the NNBP energies have an error smaller than 0.1 kcal/mol. Our optimization algorithm converges to essentially the same solution when starting from various initial conditions because the error bars of the different solutions obtained for each initial condition overlap with each other. Figure S18d shows the final FDCs obtained when using the optimal values of the NNBP energies obtained for each initial condition. The different FDCs are indistinguishable and they reproduce quantitatively the experimental FDC.

## S14. Errors in the Monte Carlo optimization

There are three kinds of errors at different levels that we will denote as $\sigma_1$, $\sigma_2$, $\sigma_3$ :

1. The first error $\sigma_1$ comes from the fitting algorithm. The uncertainties of the estimated NNBP energies ($\sigma_{\varepsilon_i}$) indicate how much the error function ($E(\varepsilon_1,...,\varepsilon_{10},\varepsilon_{loop})$ see Eq. **2** main text) changes when the fitting parameters $\varepsilon_i$ are varied around the minimum. For instance, a variation of the AA/TT motif ($\delta\varepsilon_1$) around the minimum (see Figure S19) produces a larger change in the error function than a variation of the TA/AT motif ($\delta\varepsilon_{10}$). This indicates that the uncertainty of AA/TT is lower than that of TA/AT. The curvature of the minimum in each direction $\varepsilon_i$ gives the uncertainty. There is a different set of $\sigma_{\varepsilon_i}$ uncertainties for each fit (i.e. each molecule). A quantitative evaluation of the uncertainty of the NNBP parameters requires the evaluation of the $\chi^2$ function for each FDC (i.e. each fit), which is given by:

$$\chi^2(\vec{\varepsilon}) = \sum_{i=1}^{N}\left(\frac{f_i - f(x_i;\vec{\varepsilon})}{\sigma_y}\right)^2 \qquad \textbf{[18]}$$

where $N$ is the number of experimental points of the FDC; $x_i$ and $f_i$ are the position and the force measurements, respectively; $\vec{\varepsilon}$ is the vector of fitting parameters $\{\varepsilon_i\}$ $i$ = 1,...10; $f(x_i;\vec{\varepsilon})$ is the theoretically predicted FDC according to the model (see *Calculation of the equilibrium FDC* in the *Materials and Methods* section in main text); and $\sigma_y$ is the experimental error of the force measurements performed with the optical tweezers. As indicated in section S2, the resolution of the instrument is $\sigma_y = 0.1$ pN. The uncertainty of the fit parameters is given by the following expression (8):

$$\sigma_{\varepsilon_i} = \sqrt{C_{ii}} \qquad \textbf{[19]}$$

where $C_{ii}$ are the diagonal elements of the variance-covariance matrix $C_{ij}$. In a non-linear least square fit, this matrix can be obtained from $C_{ij} = 2 \cdot H_{ij}^{-1}$, where $H_{ij}^{-1}$ is the inverse of the Hessian matrix $H_{ij} = \frac{\partial^2 \chi^2(\vec{\varepsilon}_m)}{\partial\varepsilon_i \partial\varepsilon_j}$ of $\chi^2(\vec{\varepsilon})$ evaluated at point $\vec{\varepsilon}_m$ that minimizes the error. Note that the error function and the $\chi^2$ function are related by a constant factor, $\chi^2(\vec{\varepsilon}) = \left(N/\sigma_y^2\right) \cdot E(\vec{\varepsilon})$, so their Hessians are related by one constant factor, as well. The calculation of $\sigma_{\varepsilon_i}$ is quite straightforward and it gives values between 0.003-0.015 kcal/mol. These values represent the first type of error that we call $\sigma_1$. Note that the Hessian matrix evaluated at the minima found with the heat-quench algorithm is very similar to the Hessian matrix evaluated at the minimum, which means that the curvature is almost the same in all heat-quench minima. Therefore the error of the fit $\sigma_1$ takes the same value within a region of ±0.1 kcal/mol.

2. The second error comes from the dispersion of the heat-quench minima. As we saw previously, there are several minima corresponding to different possible solutions (each solution being a set of 10 NNBP energies) for the same molecule. The values of the NNBP energies corresponding to the different solutions are Gaussian distributed (see Figure S10c) and the average standard deviation is about 0.05 kcal/mol. All these considerations result in a second typical error $\sigma_2$ = 0.05 kcal/mol.
3. Finally, the third error corresponds to the molecular heterogeneity intrinsic to single molecule experiments. Such heterogeneity results in a variability of solutions among different molecules. Indeed, the FDCs of the molecules are never identical and this variability leads to differences in the values of the NNBP energies. This variability is the major source of error in the estimation of our results. The error bars in Figs. 2d,e and 3 (main text) indicate the standard error of the mean, which is around 0.1 kcal/mol on average. This is what finally determines the statistical error of our analysis, $\sigma_3$ = 0.1 kcal/mol.

Since the major source of errors is the variability of the results from molecule to molecule, we simply report this last error in the manuscript. Because $\sigma_3 > \sigma_2 > \sigma_1$ we can safely conclude that the propagation of the errors of the heat-quench algorithm will not increase the final value of the error bar.

## Figures

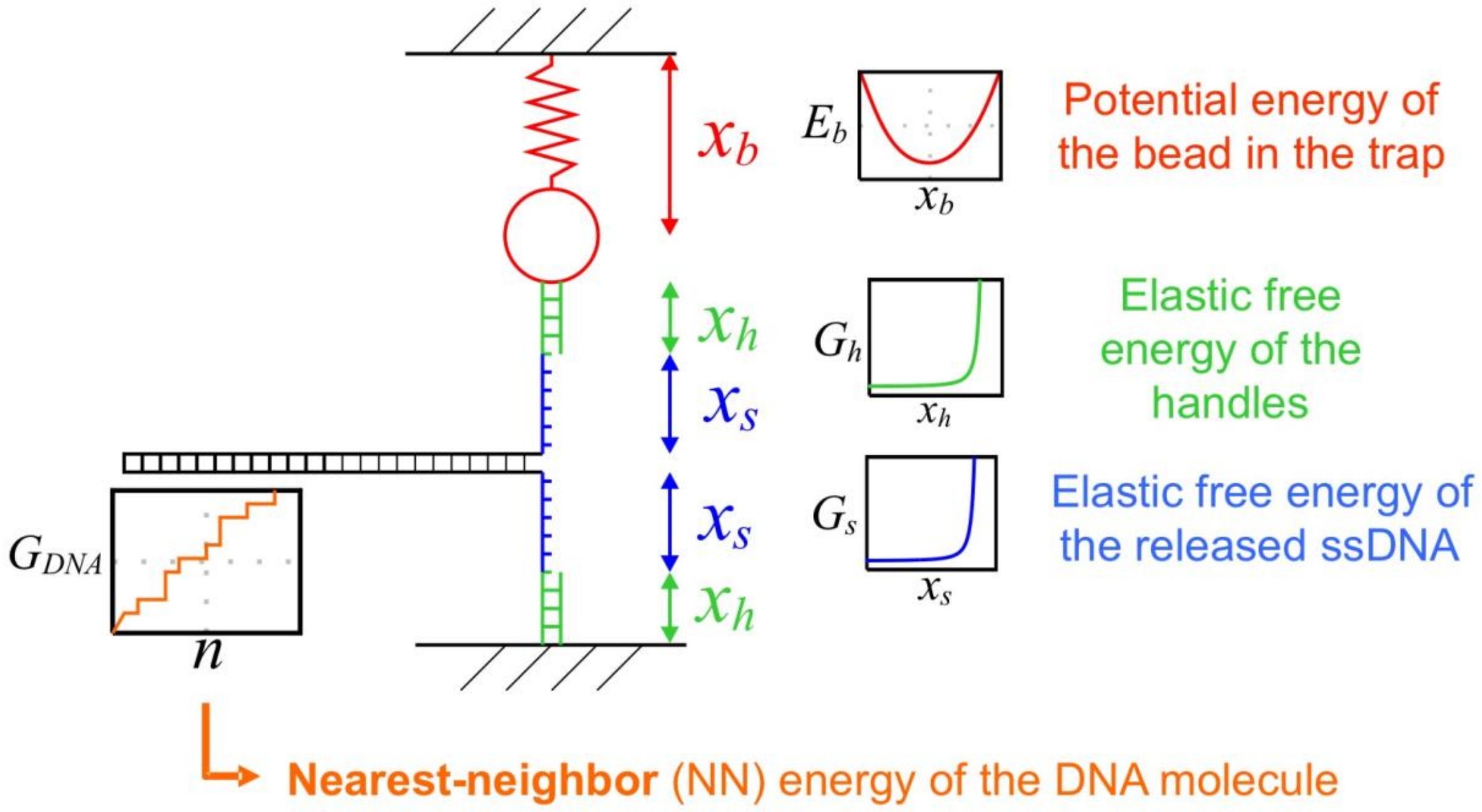

**Figure S1.** Modeling of the experimental setup. Each element is represented using a different color. A sketch of the free energy contribution is also shown for each element.

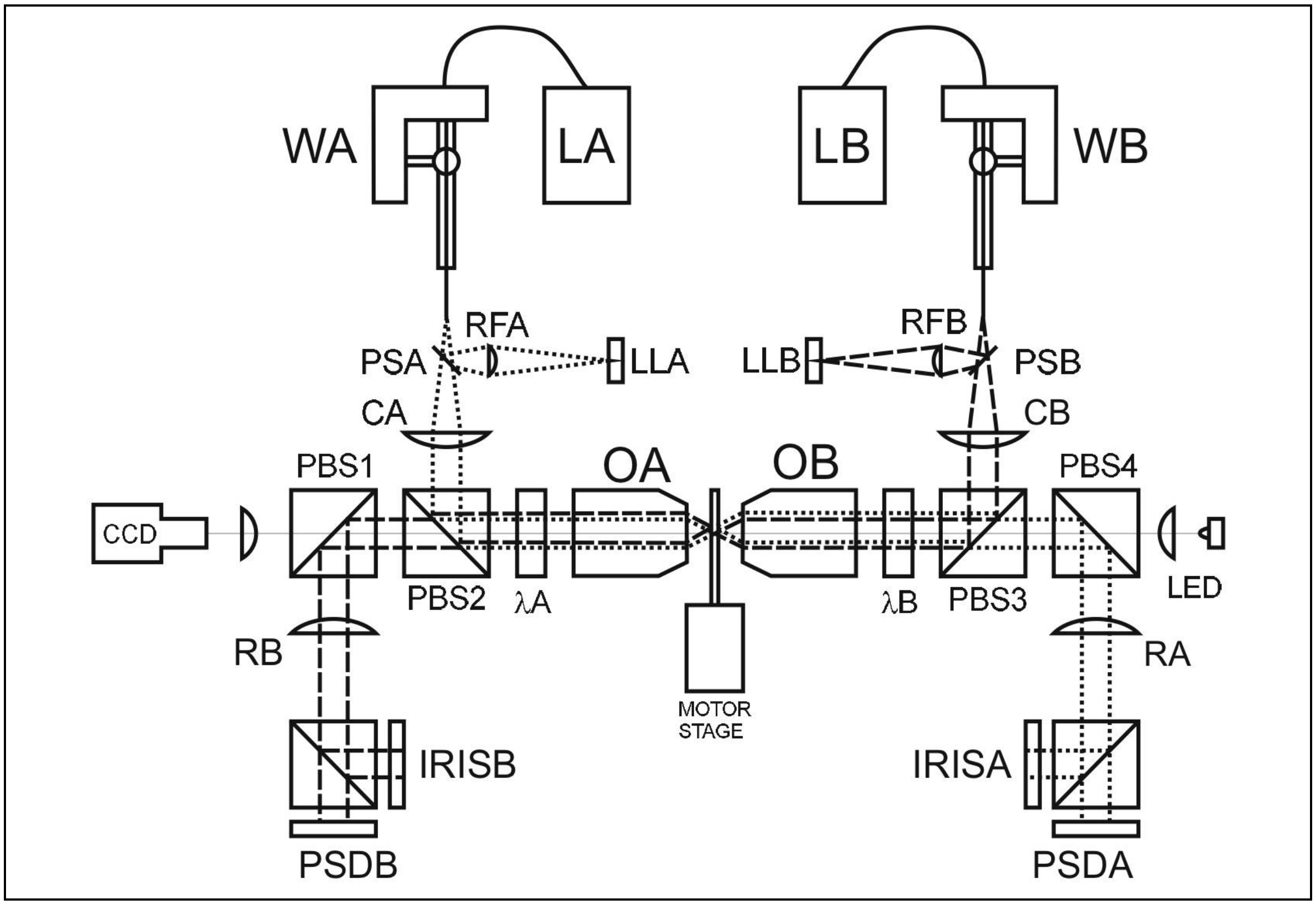

**Figure S2.** Experimental setup. The configuration is symmetric for each laser. Two fiber-coupled diodes lasers, Lumix LU845-200mW (LA & LB) feed power to twin fiber wigglers (WA & WB). The optical fiber is bent at the wiggler using piezo crystals and the laser beams can be redirected at will so that they can be repositioned. Part of the light is used to form "light-lever" position detectors by using pellicle beamsplitters (PSA & PSB), refocusing lenses (RFA & RFB) and position-sensitive detectors (LLA & LLB). The remaining light is collimated as beams by using two lenses (CA & CB) and then the beams are introduced into the optical axis by using polarizing beamsplitters (PBS2, PBS3). Two λ/4 plates produce circular polarization before the beams are focused by two Olympus 60x water-immersion microscope objectives with NA=1.2 (OA & OB). The exiting light is collected by the opposite objective (OB & OA), returned to linear (orthogonal) polarization by the λ/4 plates (λB & λA) and redirected to the (OSI Optoelectronics, DL-10) position-sensitive detectors (PSDA & PSDB) by combining the effect of the PBS and the relay lenses (RA & RB). The PSDs detect transverse (X, Y) forces while a so-called "Iris Sensor" detects changes in the axial light-momentum flux to infer the Z-axis force (2). A blue LED and a CCD camera are used to form a microscope to view the pipette and beads. The fluidics chamber is constructed from coverslips and Nescofilm gaskets (4) and its position is controlled by a motorized XYZ stage.

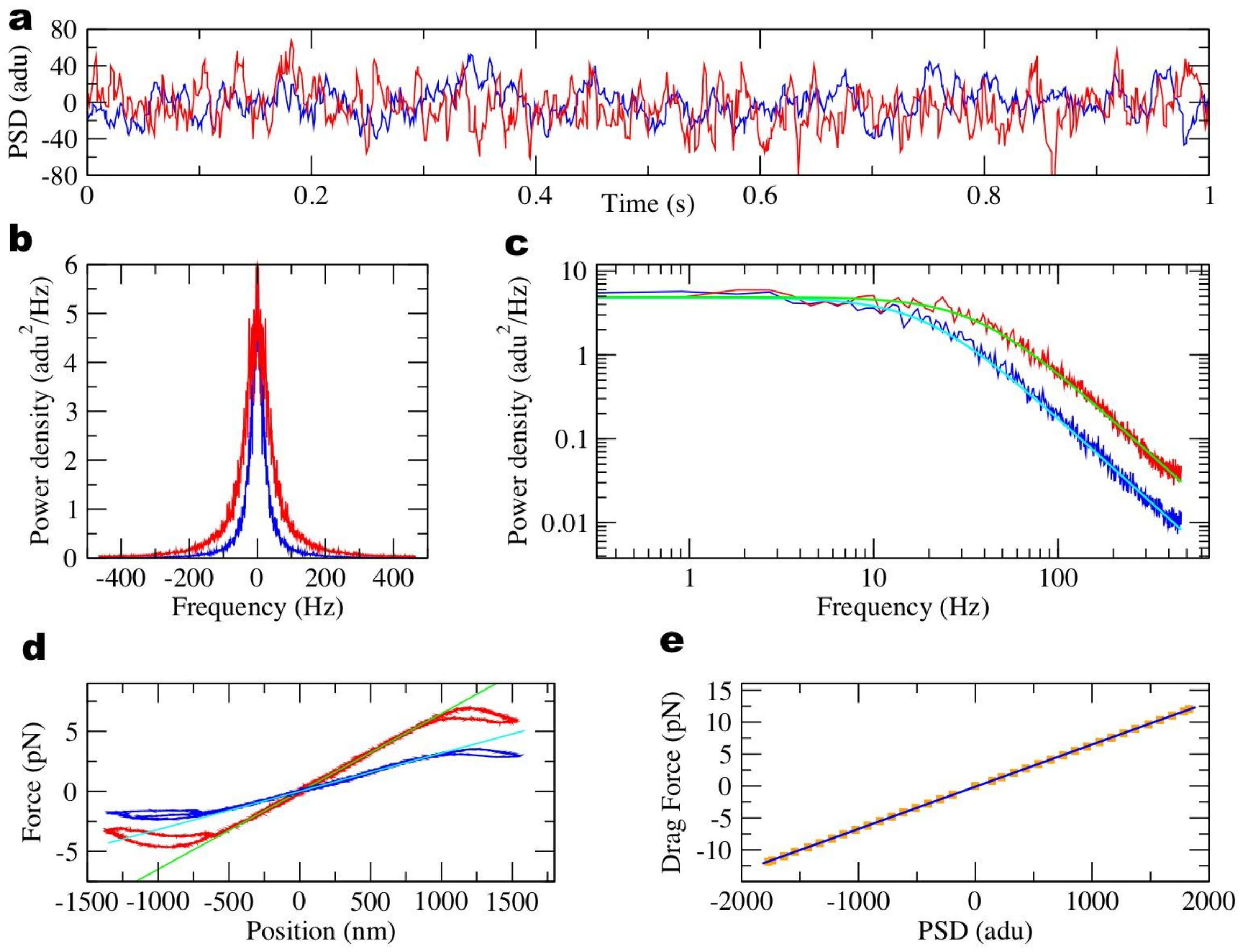

**Figure S3.** Calibration of force. **(a)** 1 second window of force reading (raw data from PSDs) versus time of two trapped particles. The units of PSD are adu (analog to digital units). The red trace shows the force reading for a particle trapped in a stiff trap and the blue trace, for a soft trap. **(b)** Noise power density for the previous recordings. Same color code than in panel a. **(c)** Noise power density in log-log scale and Lorentzian fits. Green curve shows the Lorentzian fit for the stiffer trap (red spectrum) and cyan curve shows the Lorentzian fit for the softer trap (blue spectrum). **(d)** Force vs. elongation of a particle stuck at the tip of the micropipette. The slope of the linear region of the red curve is the trap stiffness (slope of green curve). Analogous curves for the softer trap (blue and cyan curves). **(e)** Stokes law calibration. Orange dots show the experimental measurements and the blue curve depicts the linear fit. The drag force ($y$-axis) is obtained from the velocity of the motor ($v$) and using the Stokes law ($F=6\pi\eta Rv$), where the radius of the bead ($R$) and the viscosity of water ($\eta$) are known. The $x$-axis is the force in analog-digital units (adu) read from the PSD. The slope of the linear fit is the calibration factor.

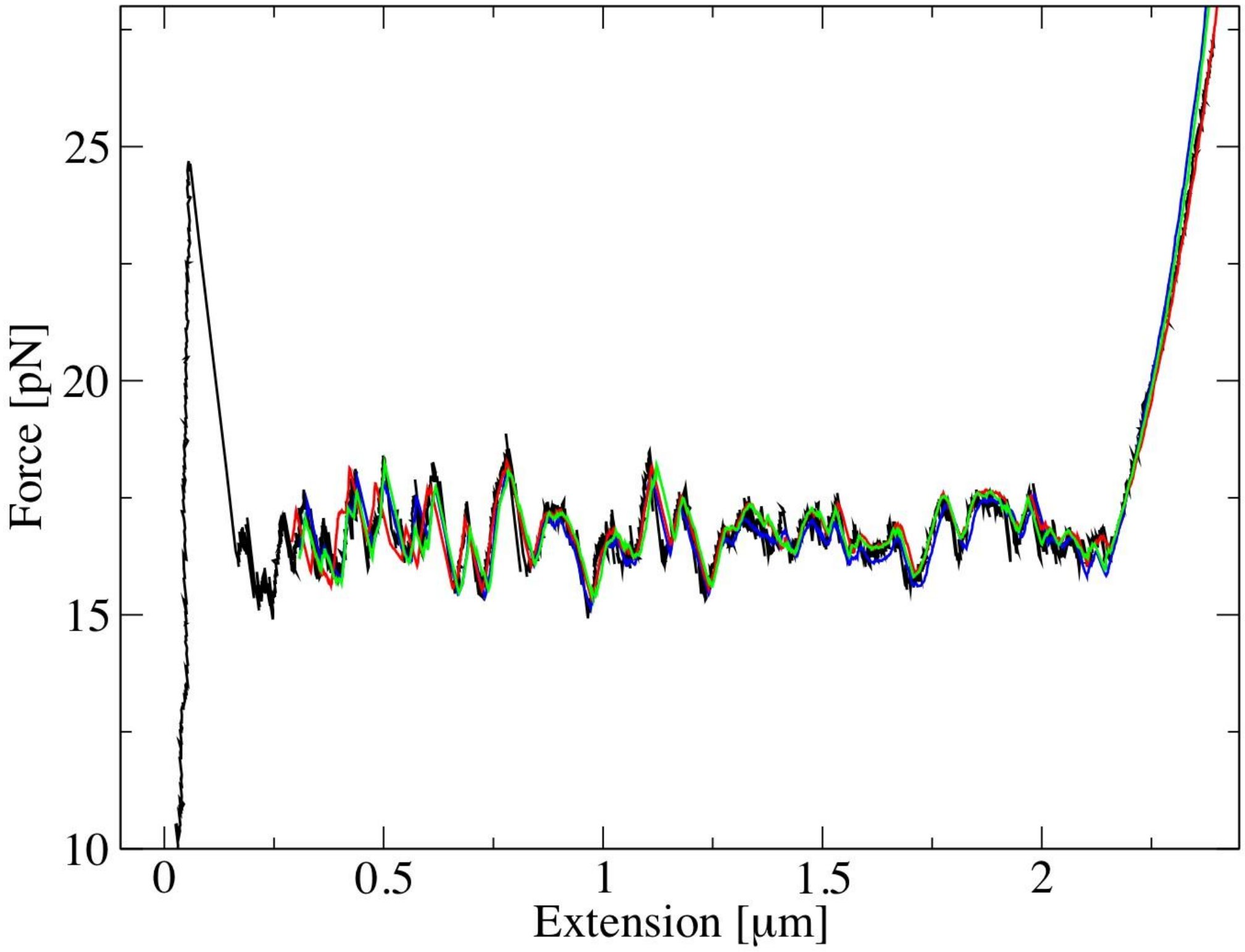

**Figure S4.** Measurements in two optical tweezers instruments. Black curve shows the unzipping data of a 2.2 kb molecule in the Berkeley setup. The data is shown as Force vs. Extension Curve (FEC). Red, blue and green curves show the data of 3 different molecules obtained with the Barcelona instrument. The FEC has been obtained from the FDC by subtracting the force/distance compliance of the optical trap.

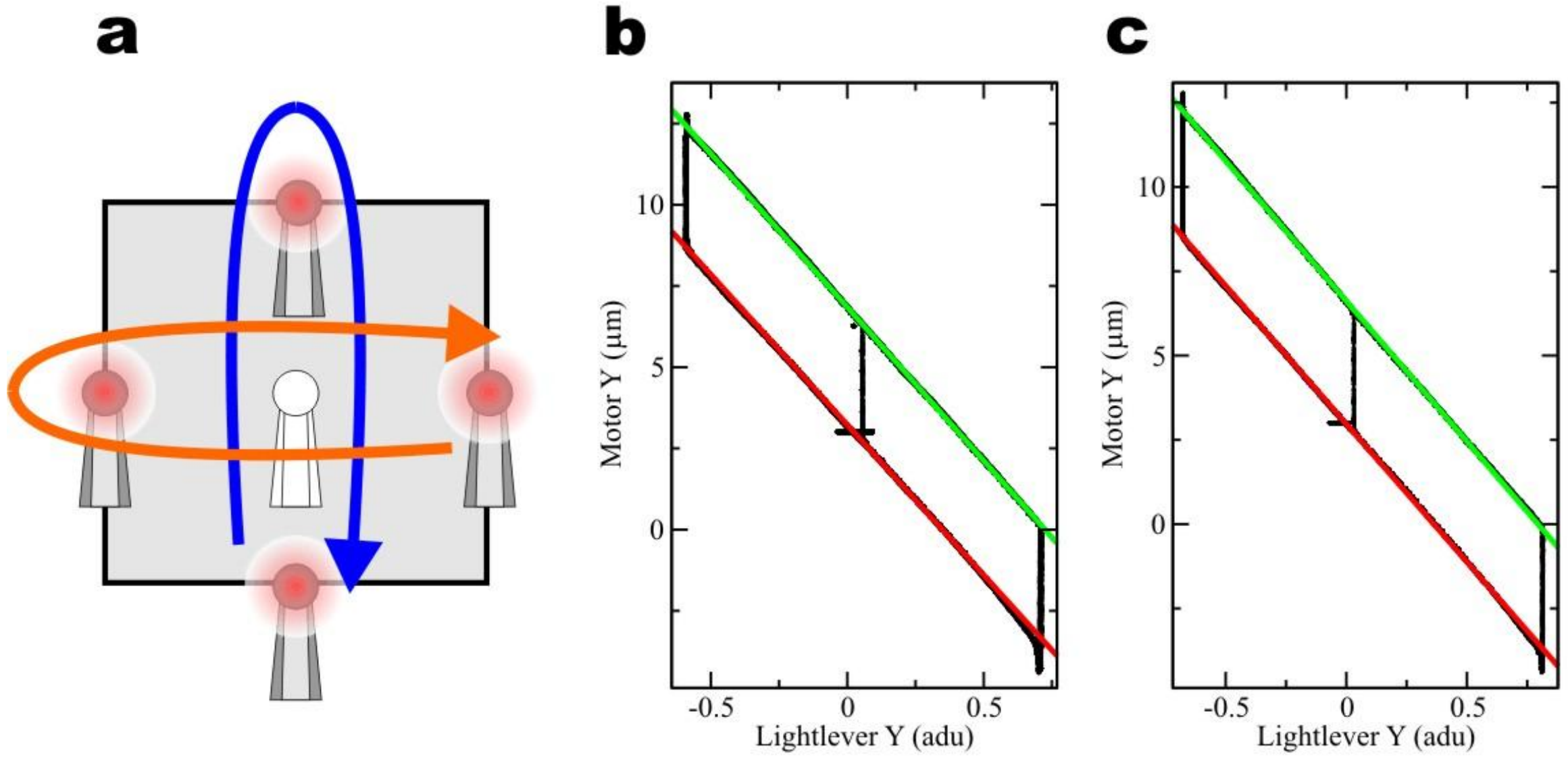

**Figure S5.** Calibration of distance. **(a)** Protocol to calibrate the light-lever position. A bead is held fixed at the tip of the micropipette. The optical trap is set to keep zero force with a force feedback algorithm operating at 4 kHz. As the pipette is gently moved the trap follows the center of the bead to maintain the preset zero force. The micropipette is moved up and down (blue arrow) to calibrate the *y* distance and left and right (orange arrow) to calibrate the *x* distance. The gray square (side length of 11 μm) shows the range of the piezos. The position of the micropipette is obtained from the shaft encoders of the motors and the position of the trap is obtained from the light-levers. **(b)** Calibration of *y* distance for trap A. The black curve is obtained moving the trap up and down. Green line shows the linear fit when the bead is moved downwards. The slope of the line is the calibration factor for the *y* distance of trap A. Red line shows the linear fit when the bead is moved upwards. Both branches (green and red) do not overlap because the motor has a backslash. During the backslash, the shaft encoder of the motor detects rotation but the gears actually do not rotate. **(c)** Calibration of *y* distance for trap B. The same procedure as in (b).

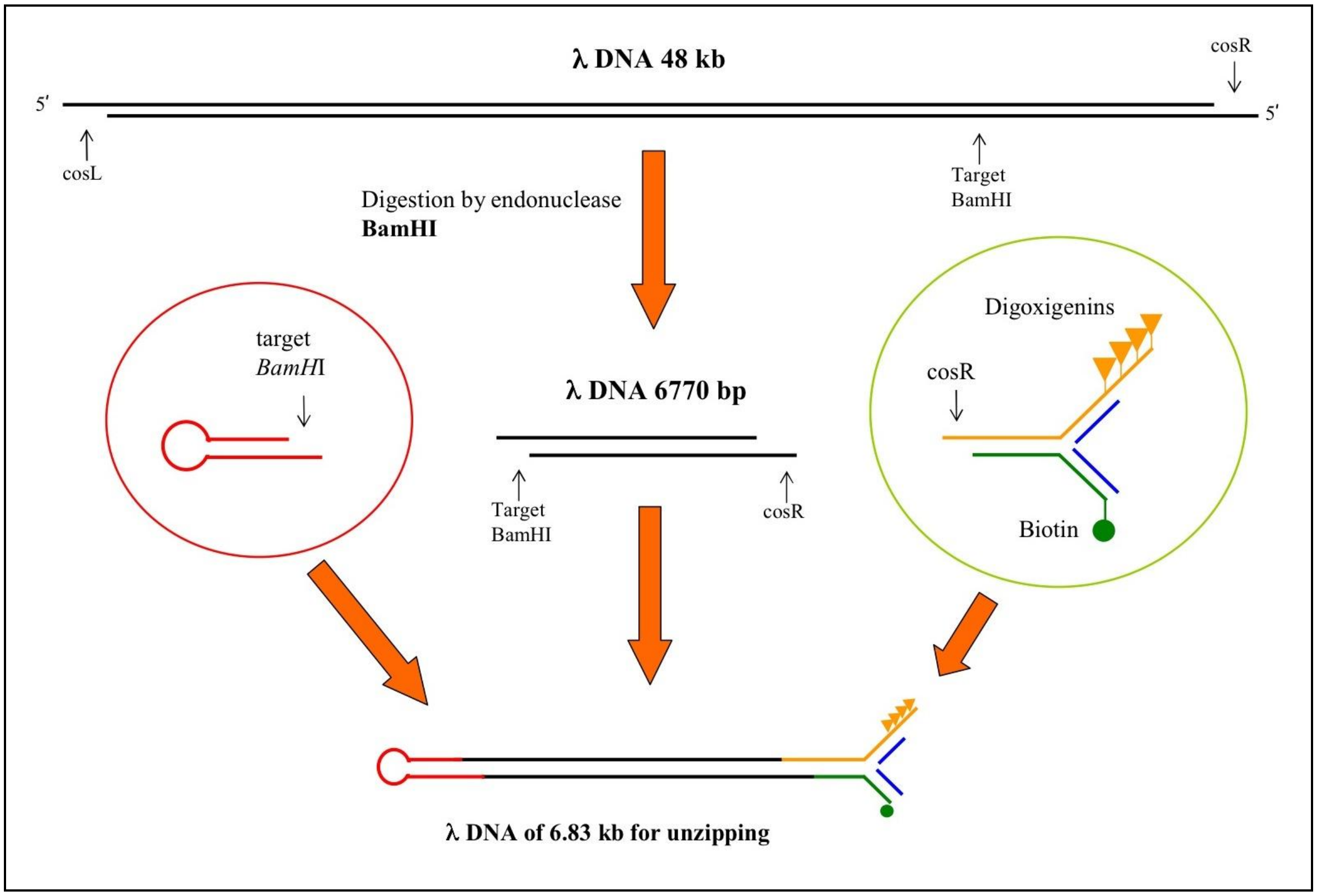

**Figure S6.** Molecular construction. A sequence of 6770 bp obtained from λ-DNA is ligated to a tetraloop and two short handles.

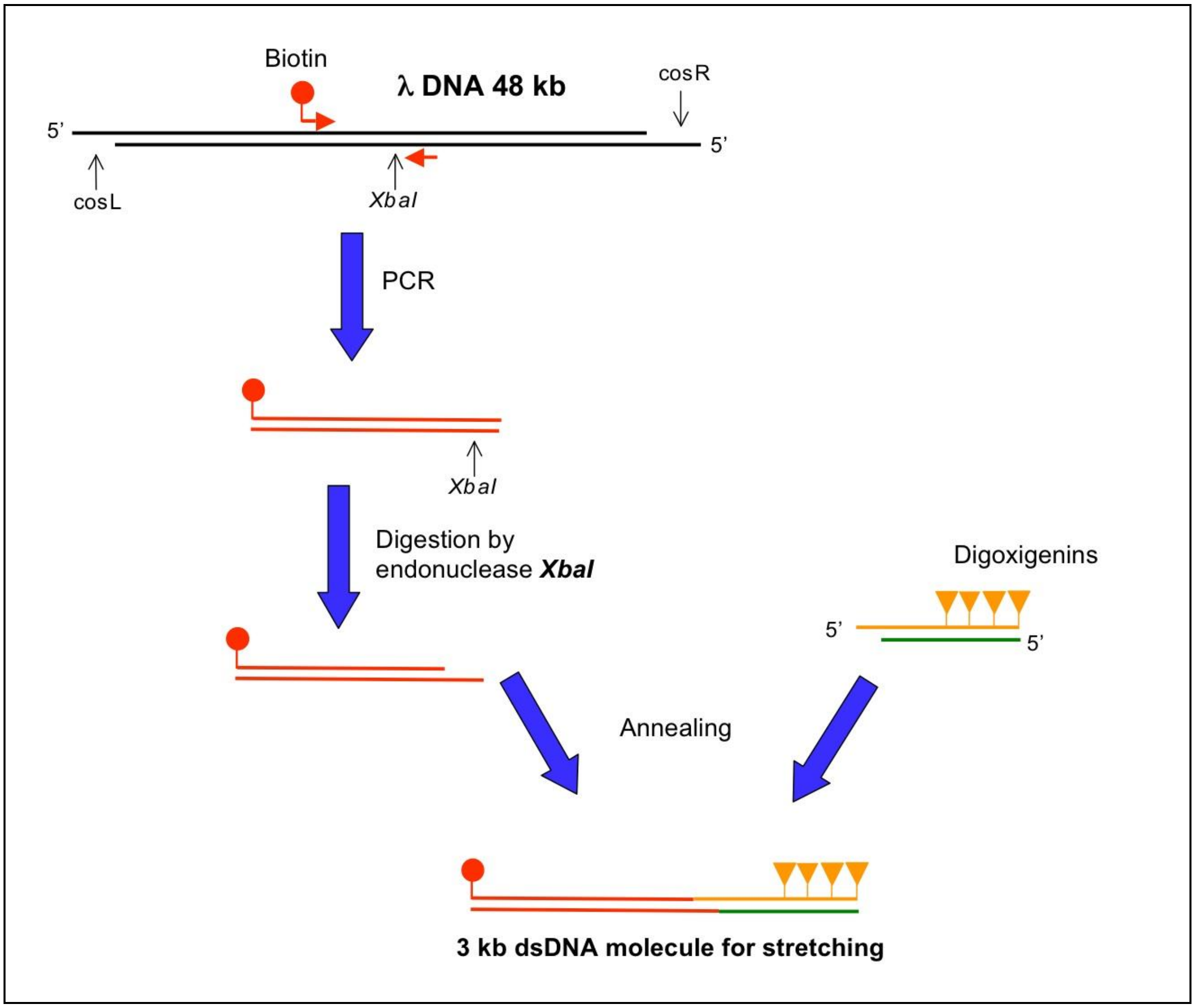

**Figure S7.** Synthesis of the 3 kb molecular construct. The denaturated ssDNA molecule can be stretched between two coated beads, since the Biotin and Digoxigenins labels are located on the same strand.

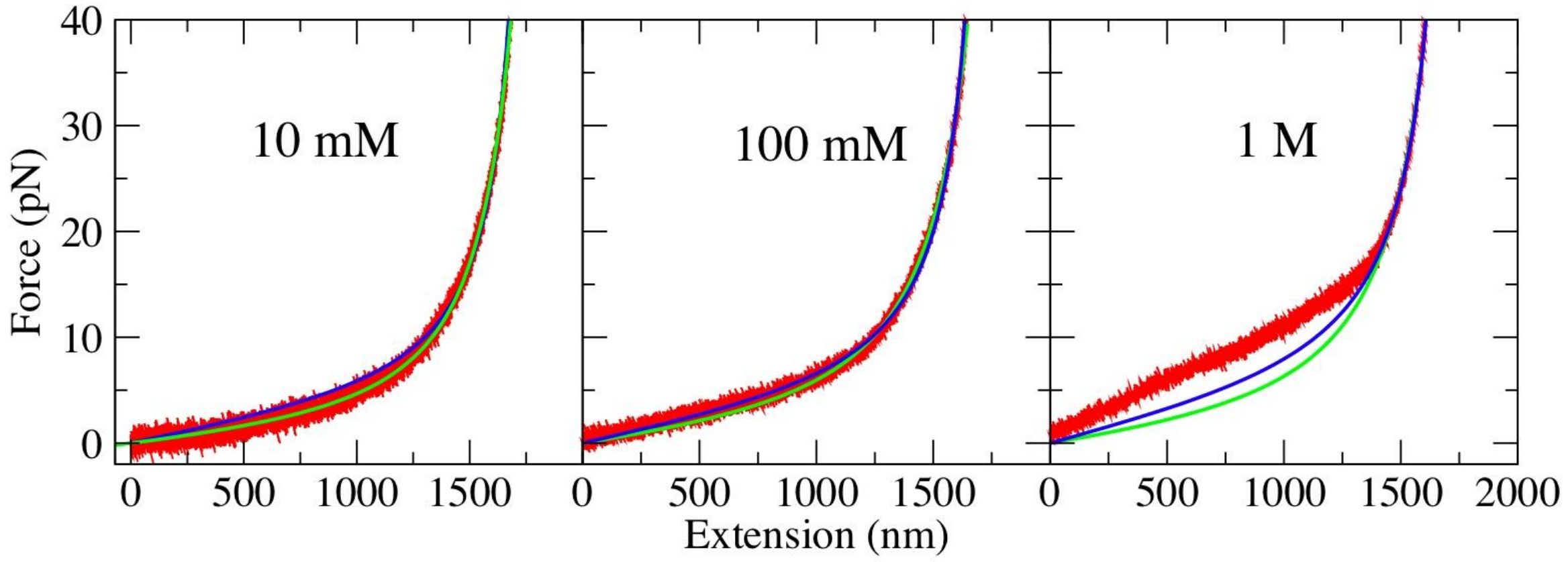

**Figure S8.** Examples of Force vs. Extension Curves (FECs) for three different salt conditions. Red curves show the experimental measured FEC (raw data). The green curves show the fit to a WLC model and blue curves show the fit to a FJC model. At 1 M NaCl, none of the two models can reproduce the experimental FEC. We assume that the observed plateau is due to the formation of secondary structure and the ideal elastic response of the ssDNA is best fit by the FJC model (blue curve).

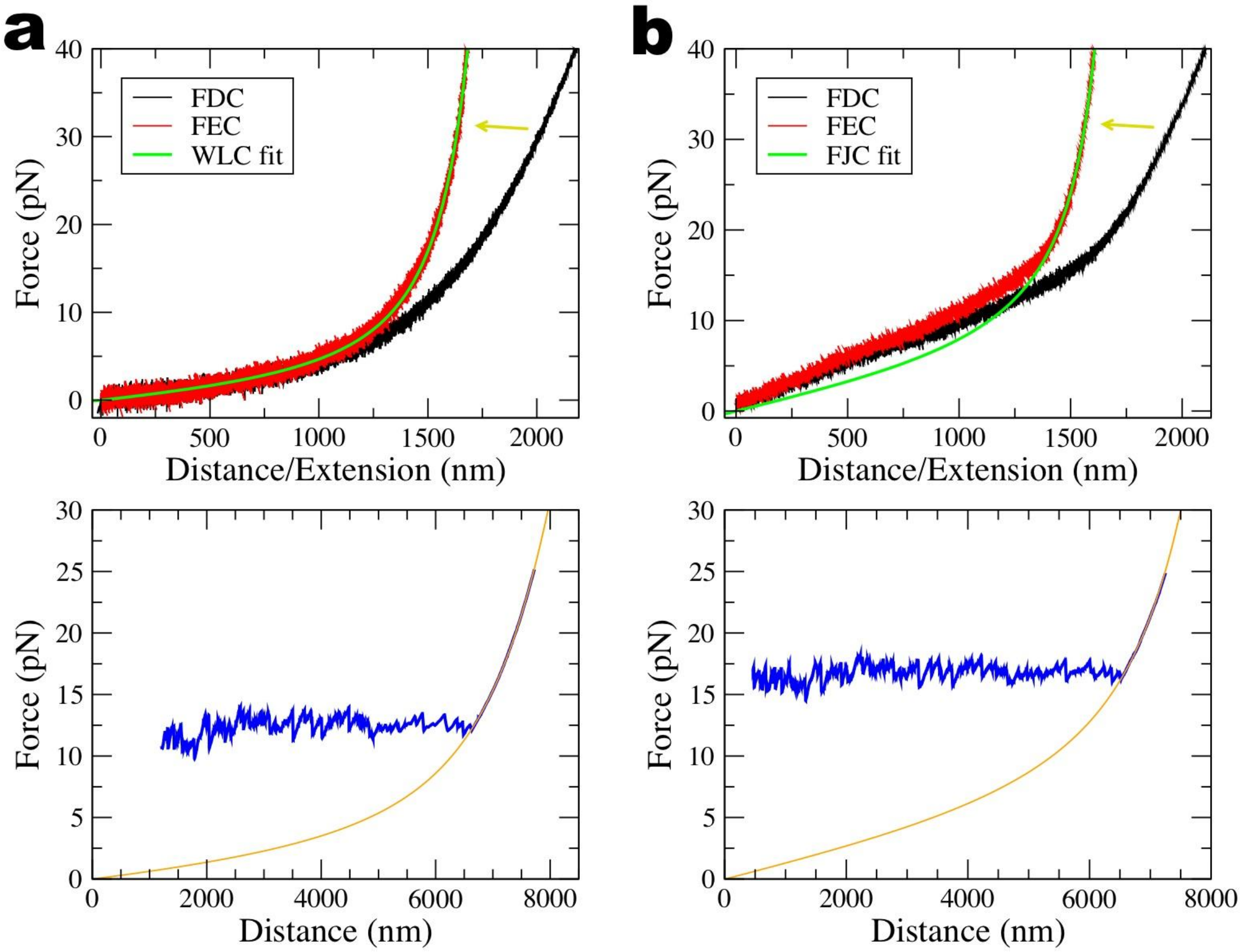

**Figure S9.** Fit of ssDNA. (a) 10 mM [NaCl]. (b) 1 M [NaCl]. Upper panels show the conversion of FDCs (black curves) into FECs (red curves) for the 3 kbp ssDNA molecule (yellow arrows indicate the direction of the conversion). The green curves show the fit of the FEC to the ideal models (WLC or FJC). Lower panels show the predicted FDC for the fully unzipped molecule (orange curve) superposed on the experimental unzipping FDC (blue curve).

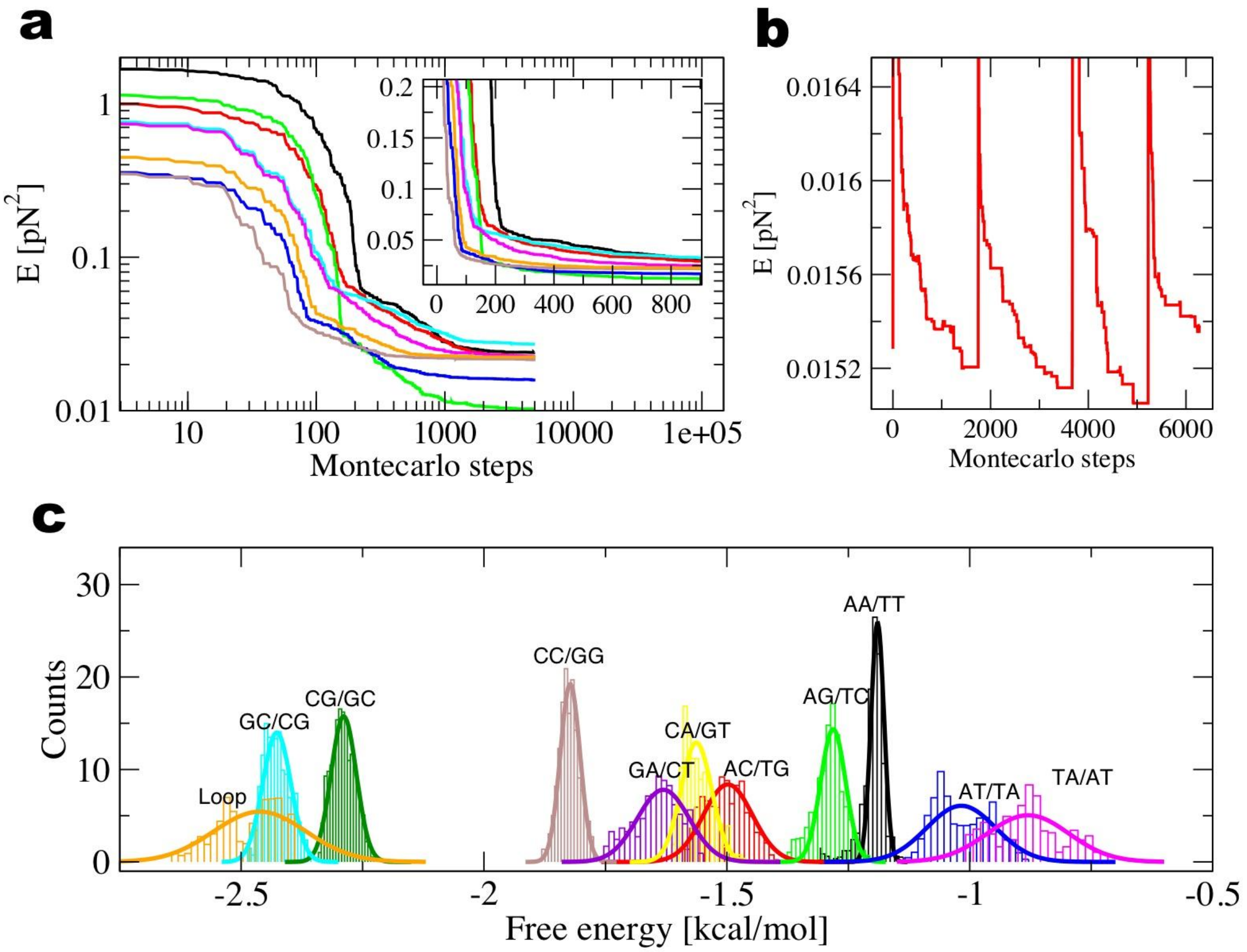

**Figure S10. (a)** Evolution of the error function of different molecules during the quenching minimization. The main figure shows a log-log plot where the mean quadratic error decreases down close to 0.01 $pN^2$. The inset figure shows a linear plot of the same evolution. **(b)** Evolution of the error during the heat-quench algorithm. **(c)** Histograms of solutions for one representative molecule obtained using the heat-quench algorithm. Each color represents one NNBP parameter and its Gaussian fit profile. Optimal solutions correspond to the most probable values of the distribution.

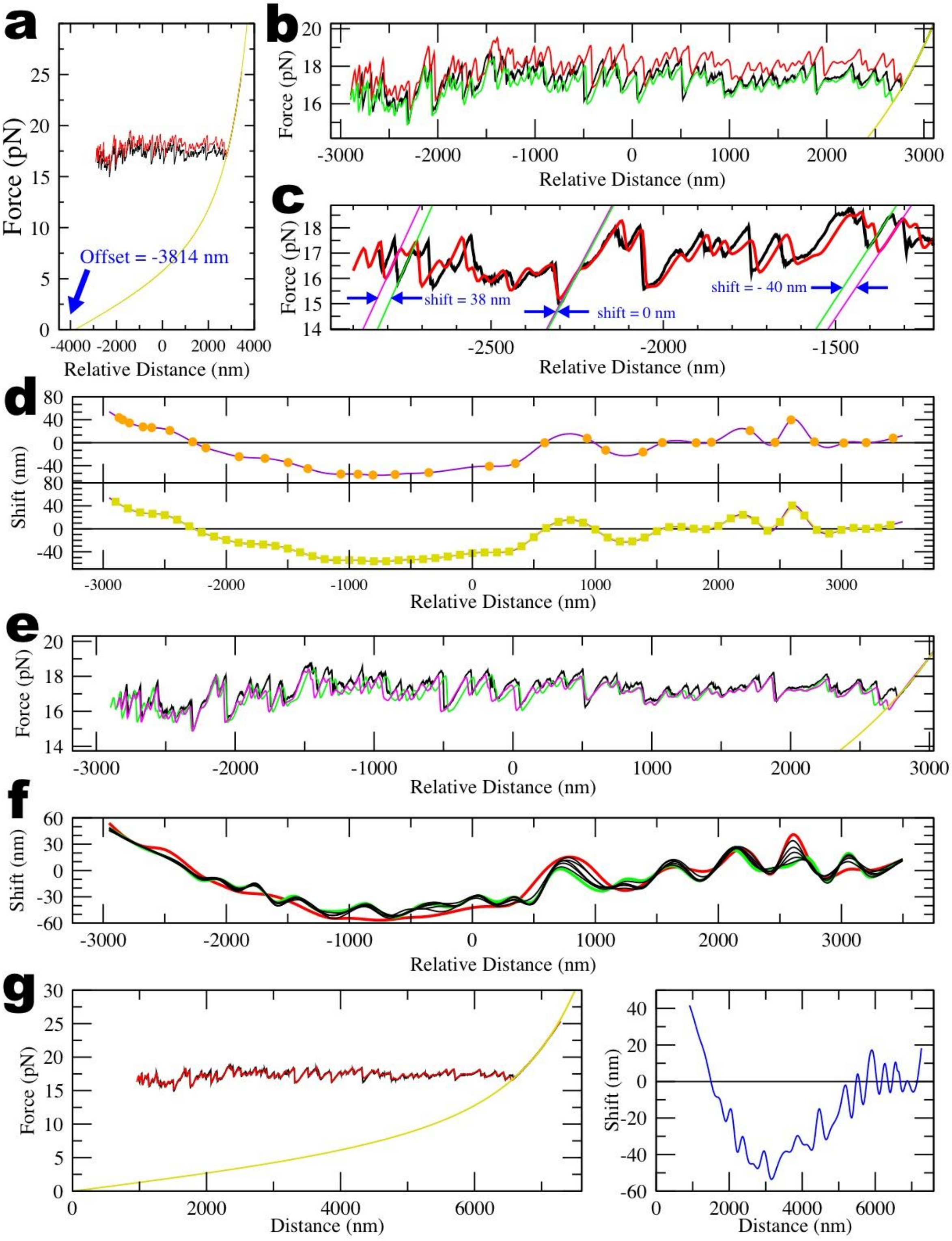

**Figure S11.** Fit of the shift function. **(a)** Step 1. **(b)** Steps 1,2 and 3. **(c)** Step 4. **(d)** Steps 4,5 and 7. **(e)** Step 6. **(f)** Step **7. (g)** Step 7.

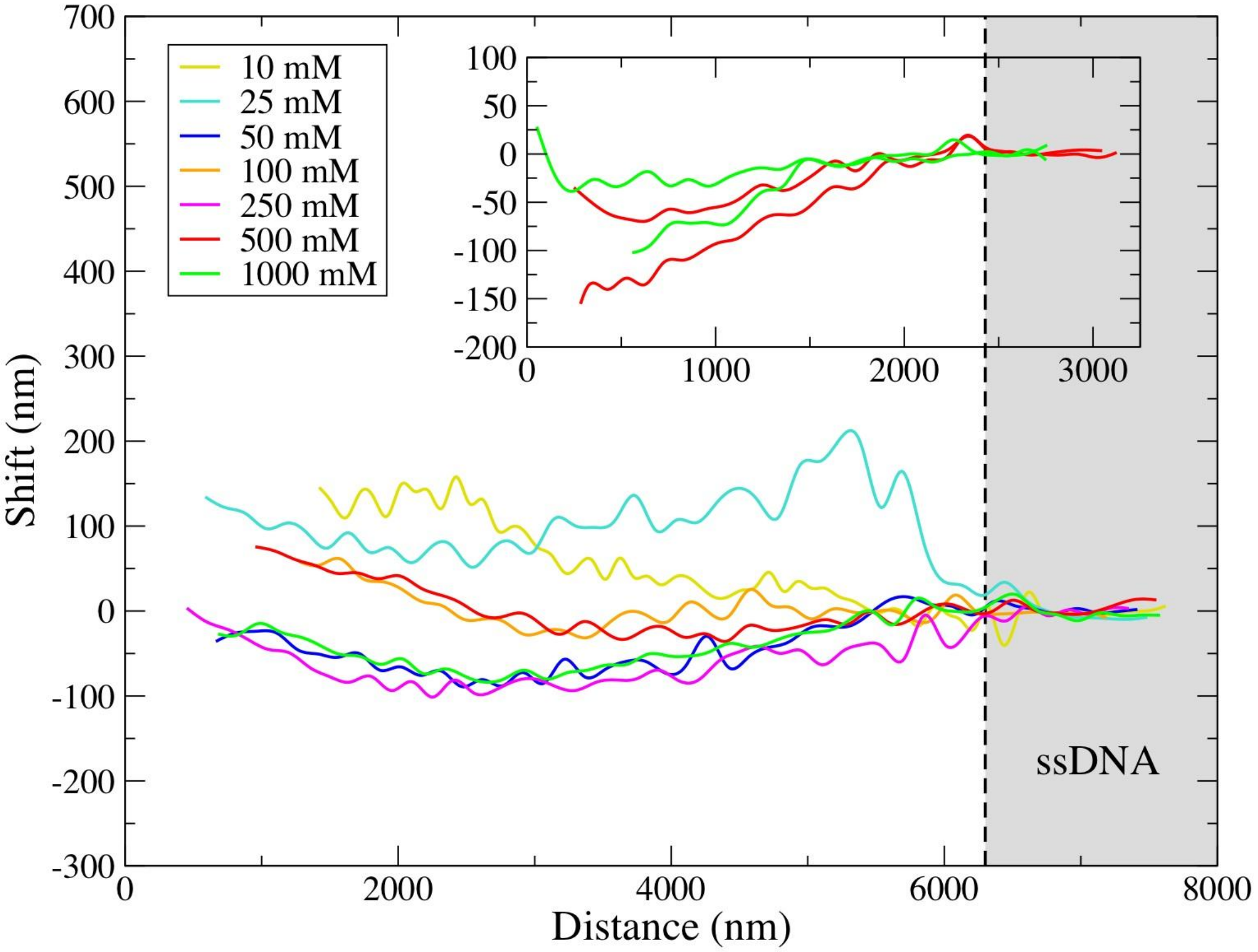

**Figure S12.** Shift function. The figure shows the shift function for some 6.8 kb molecules at different salt conditions. The inset shows the shift function for different molecules of 2.2 kbp at 500 mM NaCl (=red) and 1 M NaCl (=green). The grey shaded region corresponds to trap positions where DNA is fully unzipped. In this region, the local shift nearly vanishes.

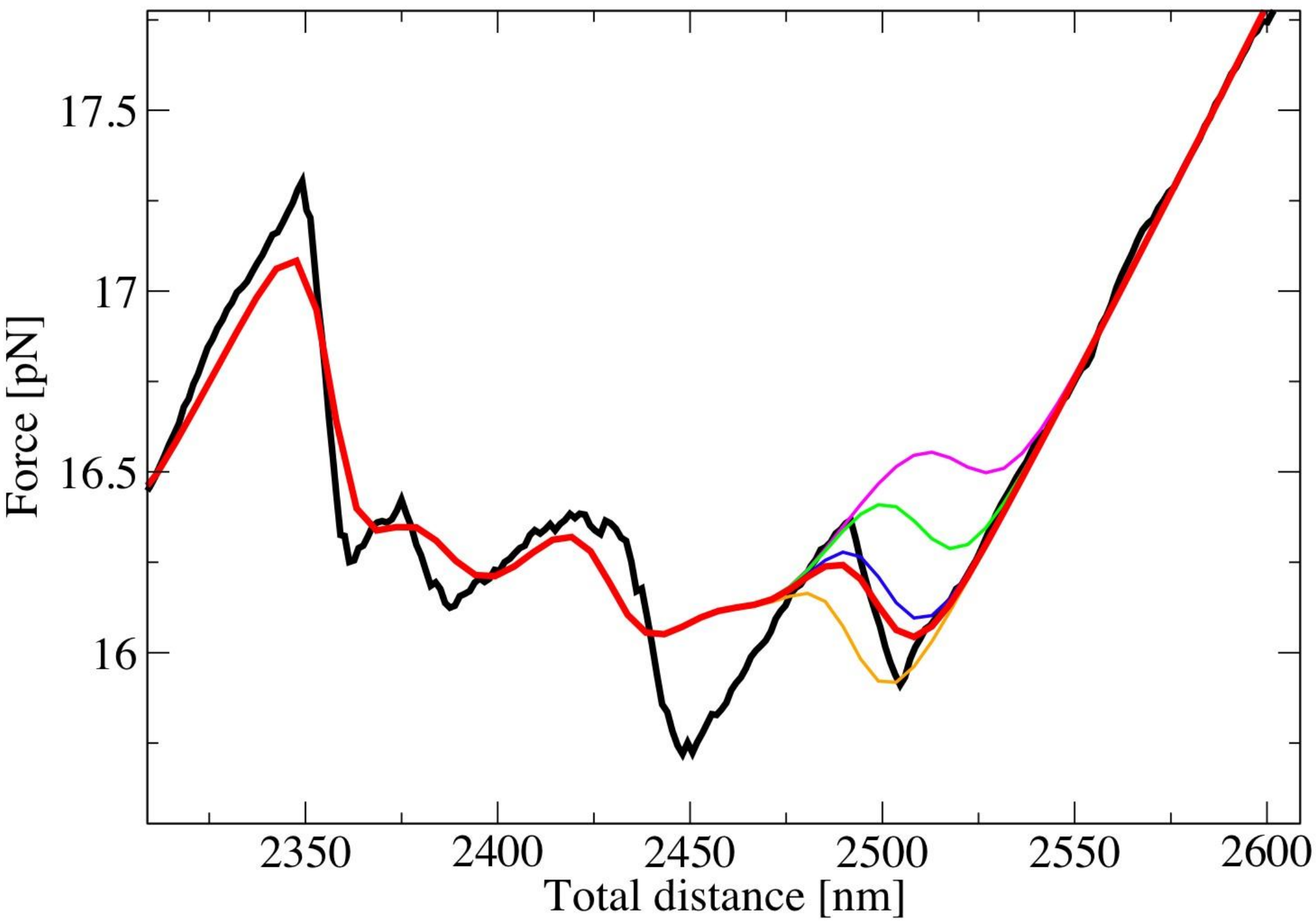

**Figure S13.** Effect of the loop contribution. The free energy of the loop modifies the shape of the theoretical FDC only at the last rip just before the elastic response of the full ssDNA is observed. The black curve is the experimental FDC. All other curves show theoretical FDCs with different values of $\varepsilon_{loop}$. Red curve, best fit with $\varepsilon_{loop}$=2.27 kcal/mol; magenta curve, $\varepsilon_{loop}$=0.0 kcal/mol; green curve, $\varepsilon_{loop}$=1.00 kcal/mol; blue curve, $\varepsilon_{loop}$=2.00 kcal/mol and orange curve, $\varepsilon_{loop}$=3.00 kcal/mol.

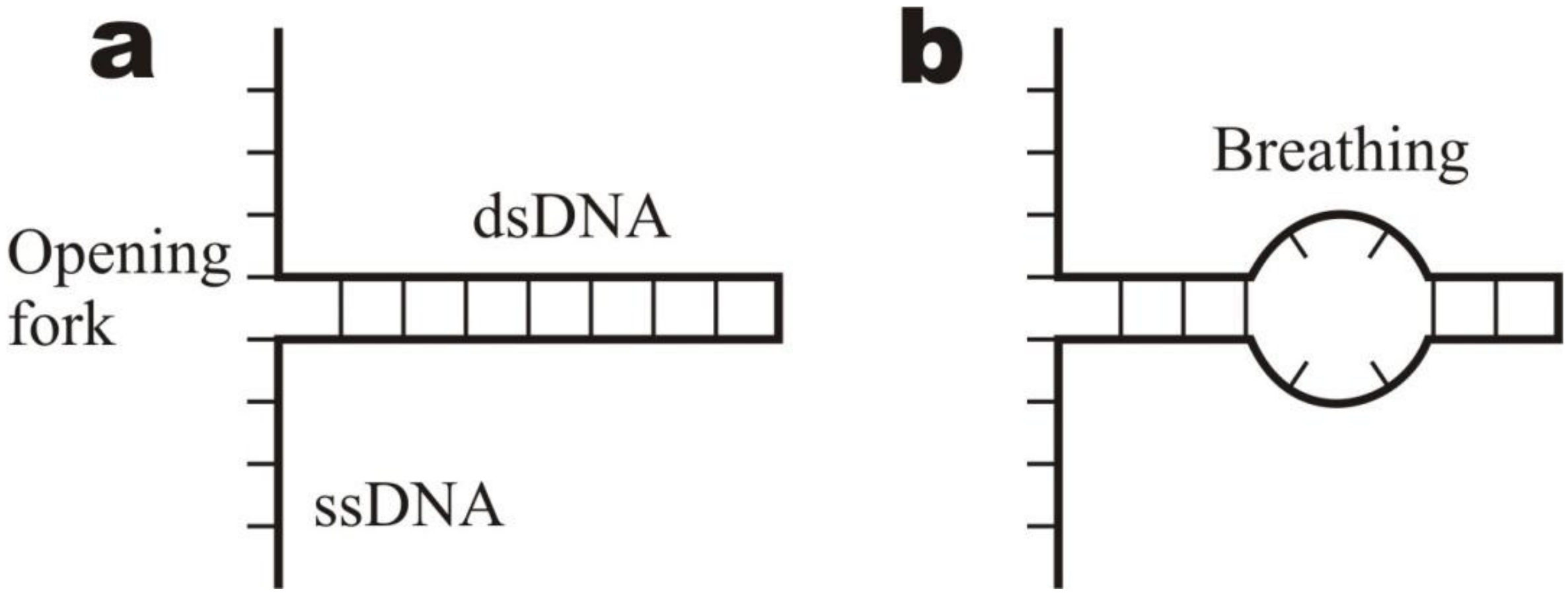

**Figure S14. (a)** Opening fork. **(b)** Breathing.

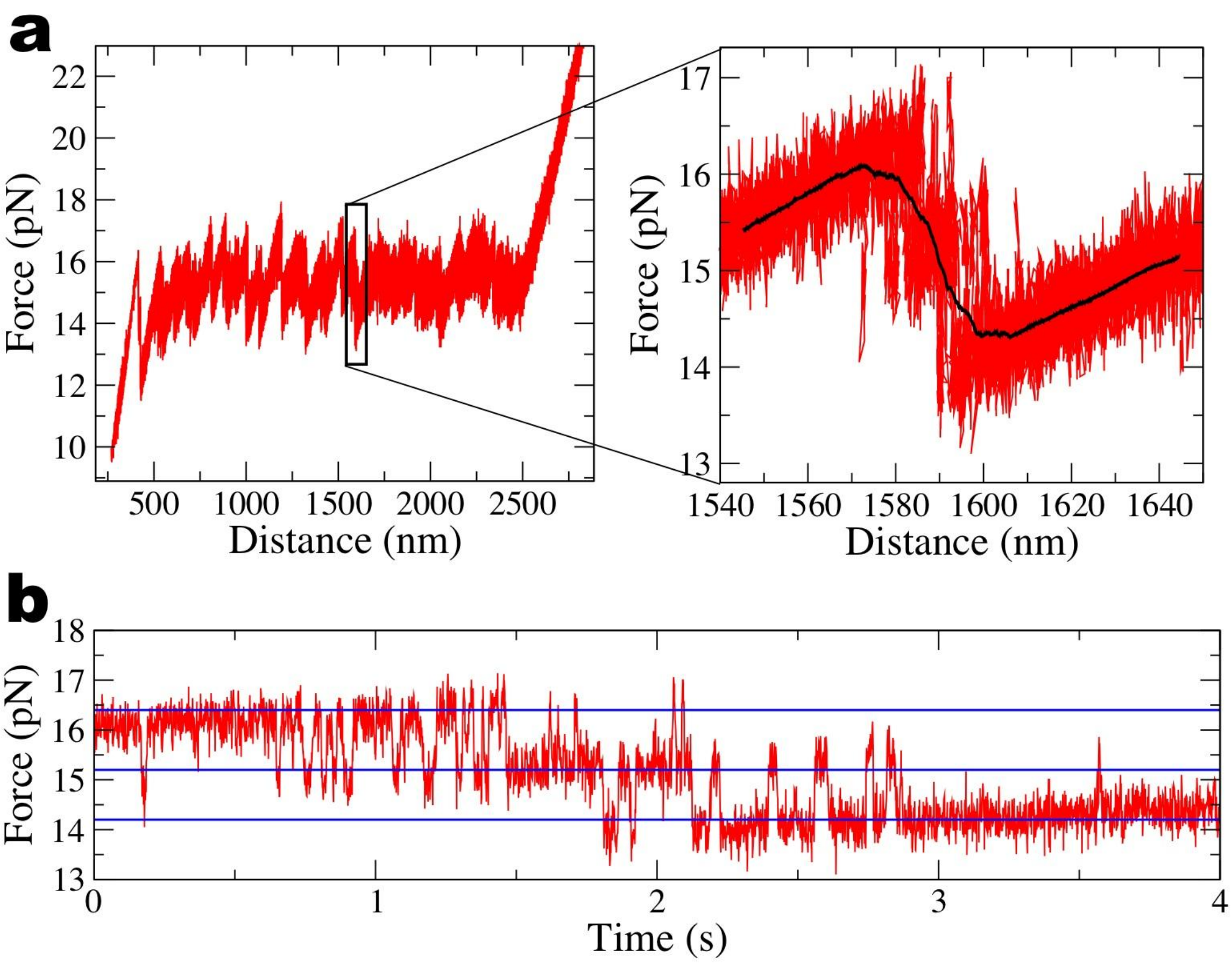

**Figure S15.** Coexistence of states. **(a)** Left panel shows the measured FDC for the 2.2 kbp sequence. Right panel shows the fragment of the FDC (framed in the left panel) where 3 states coexist. Red curve shows the raw data and black curve shows the data filtered at 1 Hz. **(b)** Red curve shows the force vs. time of the previous fragment where the transitions between these 3 states can be observed. The blue lines indicate the average forces corresponding to each of these 3 states.

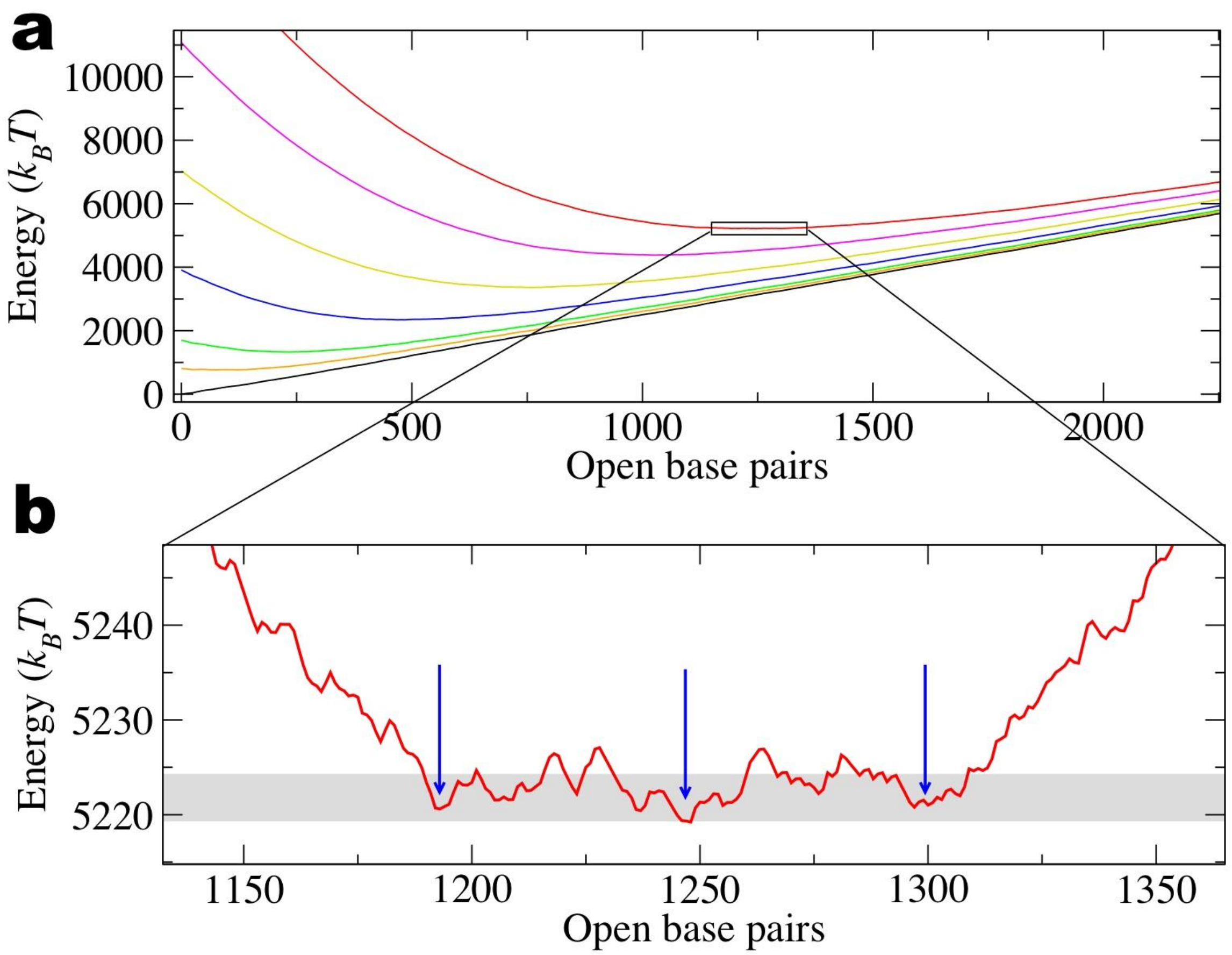

**Figure S16.** Free energy landscape for the 2.2 kb sequence at fixed distance. (a) The parabolic-like shape of the free energy landscape around the minima can be identified in a coarse grained view of what in truth is a rough landscape (see zoomed part of the landscape). Black, orange, green, blue, yellow, magenta and red curves show the free energy landscape at $x_{tot}$ = 0,350, 500, 750, 1000, 1250 and 1455 nm, respectively. (b) Zoomed region of the free energy landscape at the distance in which the 3 states of Figure S15 coexist. The blue arrows indicate the minima that correspond to these states. The highlighted gray area shows an energy range of 5 $k_BT$ .

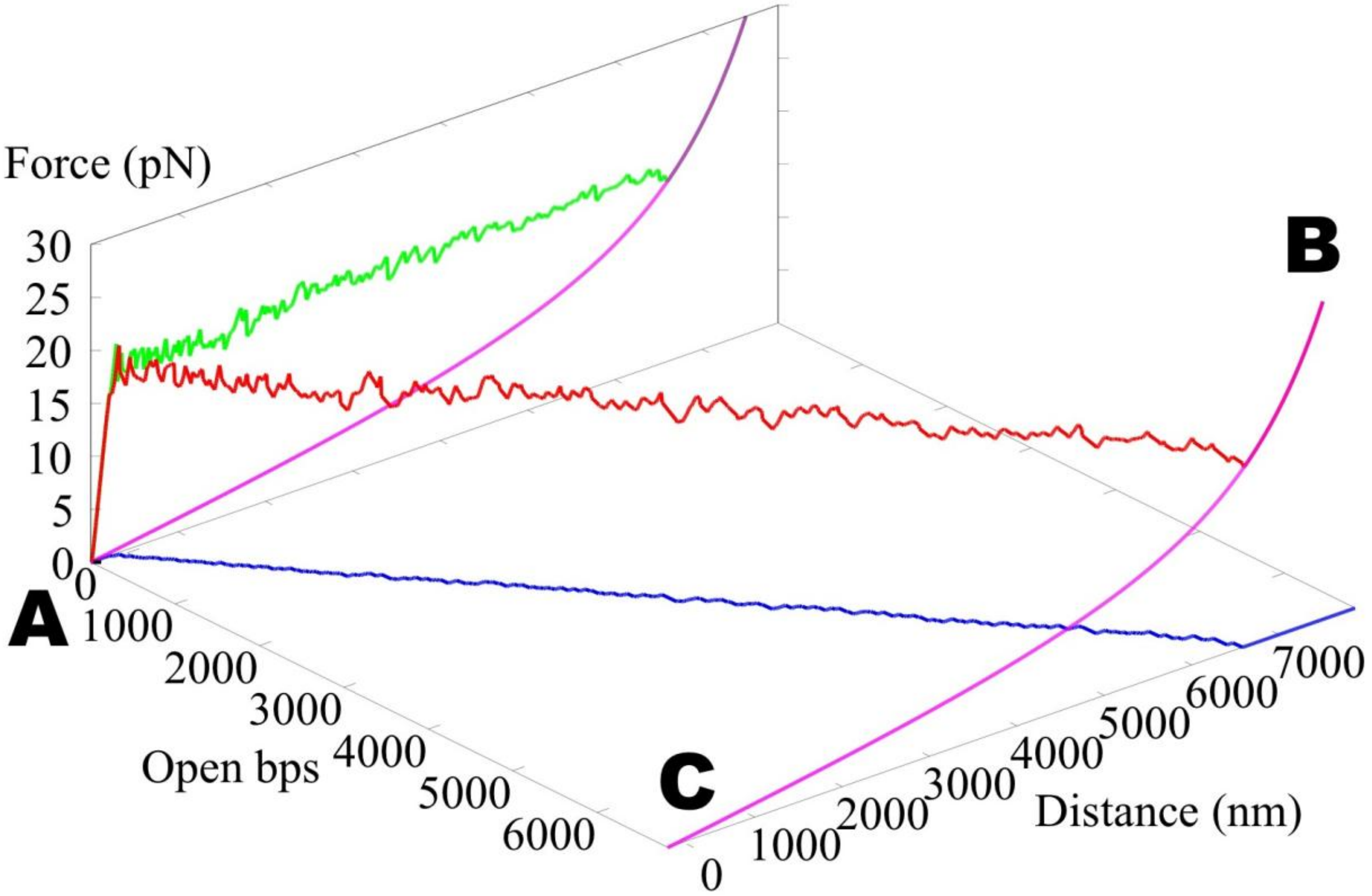

**Figure S17.** Equation of state of the 6.8 kb molecular construct. The red curve shows an equilibrium unzipping process in which the number of open bps increases as the total distance is increased. The projection of this process on the force-distance plane gives the FDC and is experimentally measured (green curve). The projection on the distance-open bp plane is shown in blue. The elastic response of the fully opened molecule is depicted in magenta (the projection in the Force-Distance plane is also depicted in magenta). **A** is the initial state in an unzipping experiment and **B** is the final state. **A** is the native state of the molecular construct. State **B** corresponds to a stretched random coil. **C** is an experimentally inaccessible state, which corresponds to a relaxed random coil.

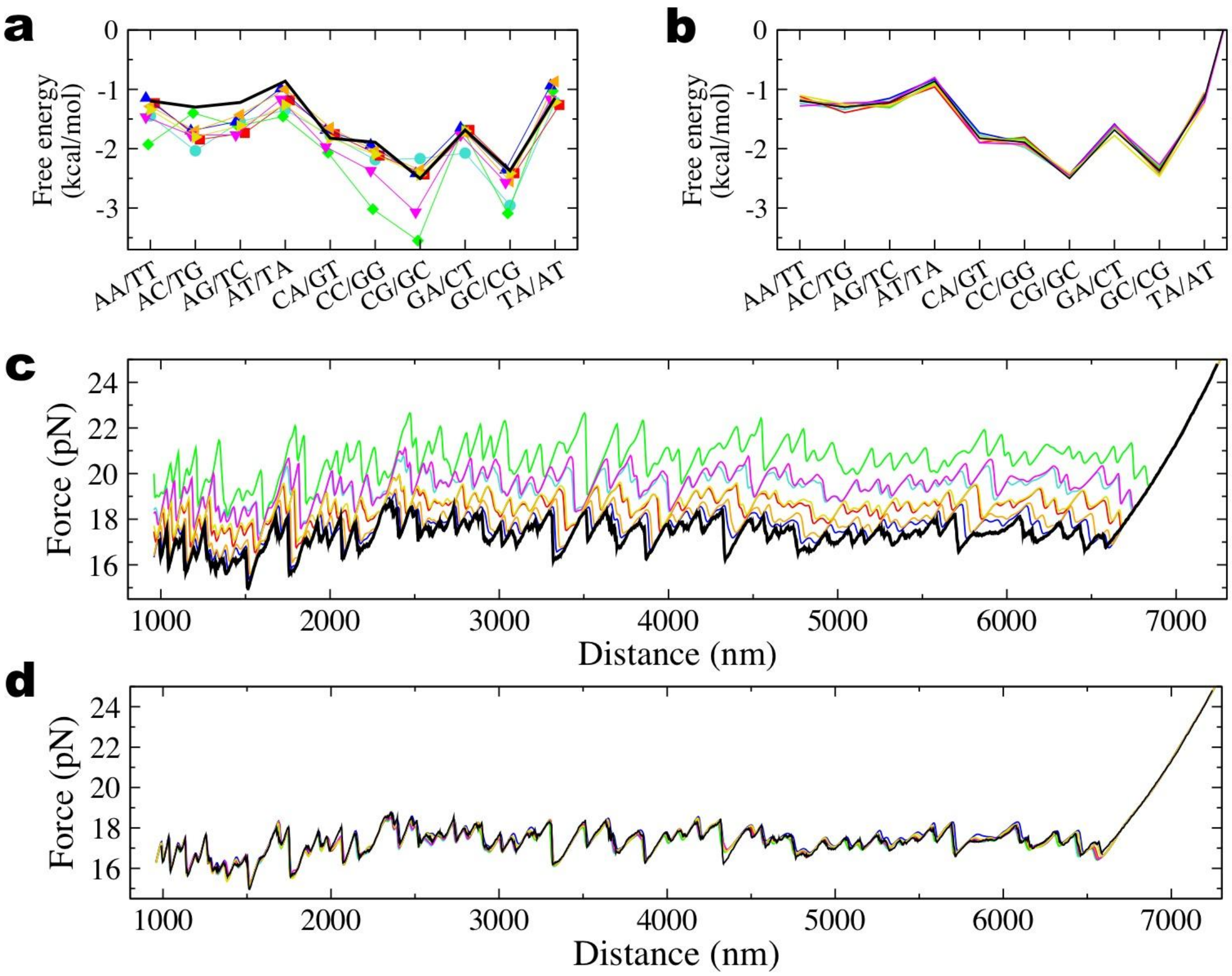

**Figure S18.** Dependence on initial conditions taken from bulk measurements (6). **(a)** Ensemble of initial conditions that we tested. Data was obtained from SantaLucia (6) and it was corrected according to temperature (298 K) and salt conditions (1 M NaCl) of our experiments. Cyan points, values of Benight; red points, values of Blake; green points, values of Breslauer; blue points, values of Gotoh; orange points, values of SantaLucia; magenta points, values of Sugimoto; yellow points, values of Vologodskii; black curve, our fit values. **(b)** Solutions found after the fitting algorithm for the different initial conditions. Same color code as in panel a. The heterogeneous ensemble of initial conditions has converged to similar values for all the NNBP energies that differ by less than 0.1 kcal/mol. **(c)** FDCs obtained using the NNBP energies from the different initial conditions. The color code is the same as in panel a. The black curve shows our fit FDC. Some of the initial conditions (Gotoh and Santalucia) are compatible with the experimental FDC. **(d)** FDCs obtained using the optimal NNBP energies obtained for each initial condition. Experimental and optimal FDCs differ by less than 0.1 pN throughout the molecule.

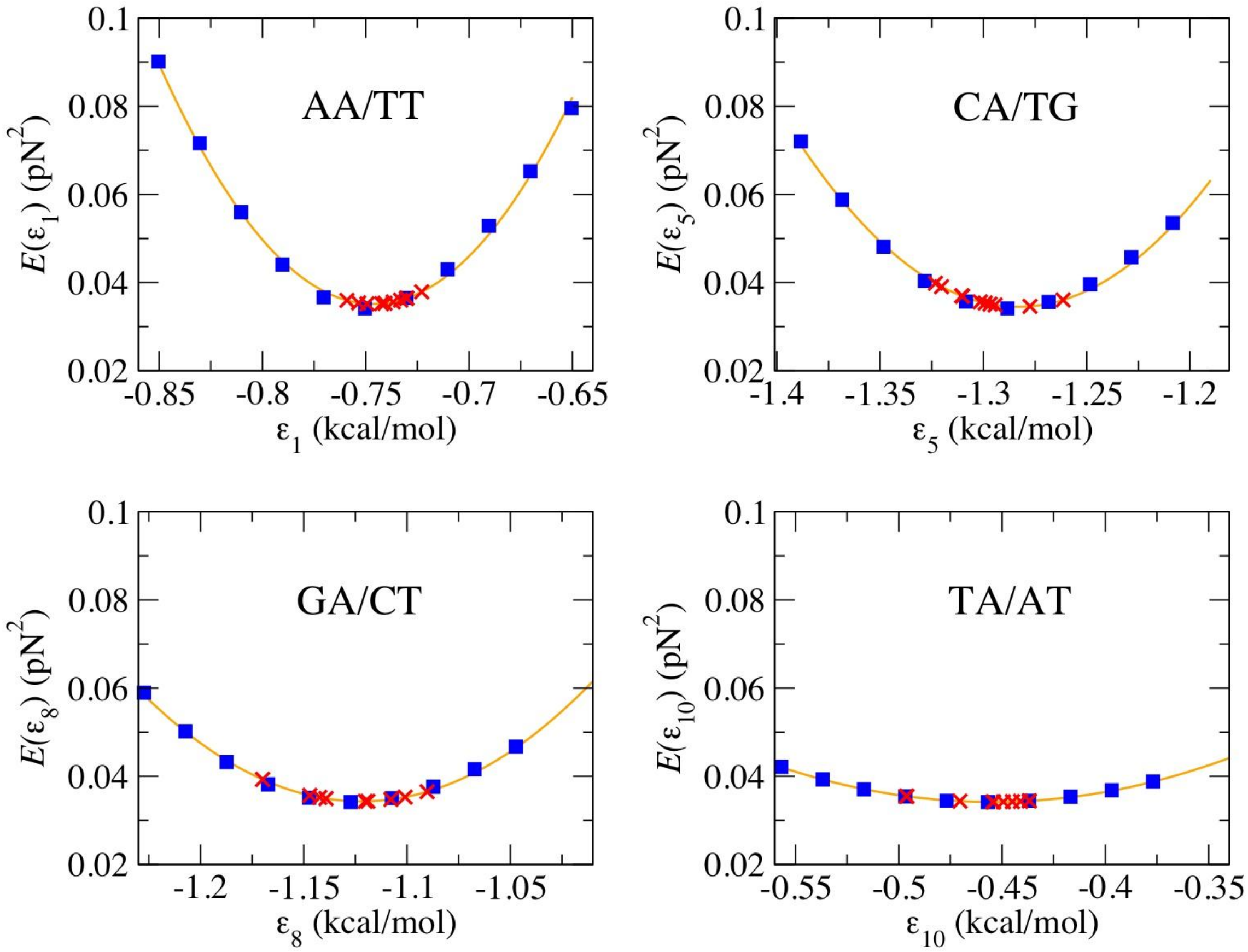

**Figure S19.** Error function ( $E(\varepsilon_1,...,\varepsilon_{10})$ see Eq. 2 main text) around the minimum for small variations of some NNBP energies. Blue dots show the error function evaluated at different values of $\varepsilon_i$. Orange curves show the quadratic fits around the minimum according to $E(\varepsilon)=c/2\cdot(\varepsilon-\varepsilon_0)^2+E_0$, where $c$, $\varepsilon_0$ and $E_0$ are fitting parameters. Red crosses show the solutions found with the MC algorithm, which differ by less than 0.05 kcal/mol with respect to the minimum. Note that the values of the curvatures ($c$) of the error function are different for each NNBP parameter. The curvature allows us to estimate the error of the fitting parameters. We have checked that the curvature of the quadratic fit (i.e. the value of the parameter $c$) for each NNBP parameter coincides with the diagonal elements of the Hessian matrix, which give the curvature of the error function in the 10-dimensional space.

# Tables

**Table S1.** Initiation terms. The contributions are not mutually exclusive. It means that they are added to the total Δ$H^o$ and Δ$S^o$ for each motif of the oligo.

| **Term** | **Motif** | **Δ$H^o$ (kcal/mol)** | **Δ$S^o$ (kcal/mol/K)** |
|---|---|---|---|
| Constant contribution | All | 0.2 | -5.6 |
| A/T penalty | 5′-A…-3′<br>5′-T…-3′<br>5′-…A-3′<br>5′-…T-3′ | 2.2 | -6.9 |
| AT penalty | 5'-TA-…-3'<br>5'-…-TA-3' | -0.4 | 0.5 |

**Table S2.** Prediction of melting temperatures for the 92 oligos of ref. 7. $T_e$ is the experimental measured temperature in ref. 7. $T_{UO}$ is the Unified Oligonucleotide prediction obtained with the parameters of ref. 6. $T_u$ is the prediction with our values obtained with unzipping experiments. Temperatures given in Celsius degrees.

| | Melting Temperatures (ºC) | | | | | | | | | | | | | | |
|---|---|---|---|---|---|---|---|---|---|---|---|---|---|---|---|
| Salt | 69 mM | | | 119 mM | | | 220 mM | | | 621 mM | | | 1020 mM | | |
| Sequence 5'-…-3' | $T_e$ | $T_{UO}$ | $T_u$ | $T_e$ | $T_{UO}$ | $T_u$ | $T_e$ | $T_{UO}$ | $T_u$ | $T_e$ | $T_{UO}$ | $T_u$ | $T_e$ | $T_{UO}$ | $T_u$ |
| ATCAATCATA | 21.3 | 21.7 | 19.3 | 24.5 | 24.1 | 22.1 | 27.9 | 26.9 | 25.2 | 32.4 | 31.7 | 30.7 | 33.6 | 34.0 | 33.4 |
| TTGTAGTCAT | 24.7 | 24.5 | 20.7 | 28.2 | 26.9 | 23.2 | 31.2 | 29.6 | 26.1 | 34.8 | 34.4 | 31.0 | 36.0 | 36.7 | 33.5 |
| GAAATGAAAG | 22.1 | 23.2 | 18.6 | 25.3 | 25.5 | 21.3 | 29.1 | 28.0 | 24.4 | 33.1 | 32.5 | 29.9 | 34.4 | 34.6 | 32.6 |
| CCAACTTCTT | 29.0 | 28.3 | 24.2 | 32.1 | 30.7 | 26.7 | 35.9 | 33.4 | 29.5 | 39.6 | 38.1 | 34.5 | 40.6 | 40.4 | 36.9 |
| ATCGTCTGGA | 33.8 | 33.8 | 29.6 | 37.4 | 36.2 | 32.3 | 40.5 | 39.0 | 35.4 | 44.5 | 43.8 | 40.7 | 44.9 | 46.2 | 43.3 |
| AGCGTAAGTC | 27.4 | 33.3 | 29.5 | 31.2 | 35.6 | 31.9 | 34.6 | 38.3 | 34.7 | 39.5 | 42.9 | 39.4 | 40.3 | 45.1 | 41.8 |
| CGATCTGCGA | 39.2 | 38.8 | 35.0 | 42.3 | 41.1 | 37.6 | 45.6 | 43.7 | 40.6 | 48.4 | 48.3 | 45.7 | 49.1 | 50.5 | 48.2 |
| TGGCGAGCAC | 44.4 | 44.2 | 41.5 | 47.8 | 46.6 | 43.5 | 51.3 | 49.3 | 45.8 | 55.0 | 54.0 | 49.9 | 55.3 | 56.3 | 51.8 |
| GATGCGCTCG | 44.2 | 42.6 | 39.7 | 47.0 | 44.8 | 42.1 | 50.1 | 47.4 | 44.8 | 53.6 | 51.8 | 49.4 | 53.5 | 54.0 | 51.7 |
| GGGACCGCCT | 46.7 | 45.9 | 43.4 | 50.3 | 48.4 | 45.3 | 53.1 | 51.3 | 47.6 | 56.5 | 56.2 | 51.4 | 57.0 | 58.6 | 53.3 |
| CGTACACATGC | 40.4 | 39.5 | 37.8 | 43.5 | 41.8 | 40.0 | 46.1 | 44.4 | 42.5 | 49.6 | 49.0 | 46.8 | 49.9 | 51.2 | 48.9 |
| CCATTGCTACC | 38.0 | 37.4 | 36.4 | 41.7 | 39.8 | 38.4 | 44.5 | 42.5 | 40.7 | 47.9 | 47.3 | 44.7 | 48.9 | 49.6 | 46.6 |
| TACTAACATTAACTA | 35.3 | 38.6 | 37.3 | 40.4 | 41.2 | 39.9 | 44.1 | 44.2 | 43.0 | 49.3 | 49.4 | 48.2 | 51.1 | 51.9 | 50.8 |
| ATACTTACTGATTAG | 38.1 | 36.7 | 35.7 | 41.4 | 39.2 | 38.4 | 45.0 | 42.2 | 41.5 | 49.9 | 47.2 | 46.8 | 51.5 | 49.7 | 49.4 |
| GTACACTGTCTTATA | 41.0 | 41.7 | 40.5 | 44.8 | 44.3 | 43.1 | 48.3 | 47.2 | 46.0 | 52.9 | 52.3 | 51.2 | 54.8 | 54.8 | 53.7 |
| GTATGAGAGACTTTA | 39.9 | 41.7 | 39.4 | 44.2 | 44.2 | 42.2 | 47.9 | 47.2 | 45.5 | 53.3 | 52.3 | 51.1 | 55.4 | 54.8 | 53.9 |
| TTCTACCTATGTGAT | 40.6 | 41.6 | 40.4 | 44.6 | 44.2 | 43.1 | 48.1 | 47.3 | 46.2 | 52.3 | 52.5 | 51.6 | 53.7 | 55.1 | 54.3 |
| AGTAGTAATCACACC | 44.3 | 43.7 | 42.7 | 47.8 | 46.3 | 45.2 | 51.6 | 49.3 | 48.2 | 56.2 | 54.4 | 53.3 | 57.1 | 56.9 | 55.8 |
| ATCGTCTCGGTATAA | 45.5 | 45.6 | 43.4 | 49.4 | 48.2 | 46.4 | 52.9 | 51.2 | 49.7 | 57.4 | 56.3 | 55.6 | 58.6 | 58.9 | 58.5 |
| ACGACAGGTTTACCA | 47.8 | 50.1 | 48.0 | 51.2 | 52.7 | 50.6 | 55.5 | 55.7 | 53.7 | 59.8 | 61.0 | 58.9 | 61.3 | 63.6 | 61.5 |
| CTTTCATGTCCGCAT | 49.9 | 49.8 | 48.5 | 53.9 | 52.4 | 51.2 | 57.1 | 55.4 | 54.3 | 61.4 | 60.5 | 59.7 | 62.8 | 63.0 | 62.4 |
| TGGATGTGTGAACAC | 46.5 | 49.1 | 47.0 | 51.6 | 51.7 | 49.7 | 54.6 | 54.7 | 52.7 | 59.1 | 59.8 | 57.9 | 60.4 | 62.3 | 60.4 |
| ACCCCGCAATACATG | 51.3 | 52.2 | 51.8 | 55.2 | 54.8 | 54.1 | 58.5 | 57.8 | 56.8 | 62.4 | 63.1 | 61.4 | 62.9 | 65.6 | 63.7 |
| GCAGTGGATGTGAGA | 51.2 | 51.2 | 49.7 | 54.8 | 53.9 | 52.2 | 58.0 | 56.9 | 55.1 | 61.7 | 62.1 | 60.1 | 63.3 | 64.6 | 62.6 |
| GGTCCTTACTTGGTG | 47.8 | 48.8 | 47.2 | 51.6 | 51.4 | 49.6 | 55.1 | 54.4 | 52.3 | 59.1 | 59.5 | 57.1 | 60.3 | 62.0 | 59.4 |
| CGCCTCATGCTCATC | 52.8 | 53.5 | 52.4 | 56.7 | 56.0 | 54.9 | 60.1 | 58.9 | 57.7 | 63.6 | 64.0 | 62.7 | 65.8 | 66.5 | 65.1 |
| AAATAGCCGGGCCGC | 59.0 | 59.3 | 57.8 | 62.2 | 61.9 | 60.0 | 65.3 | 64.9 | 62.5 | 69.0 | 70.2 | 66.9 | 70.4 | 72.7 | 69.1 |
| CCAGCCAGTCTCTCC | 54.1 | 54.2 | 53.0 | 58.0 | 56.8 | 55.3 | 61.5 | 59.9 | 57.9 | 65.1 | 65.1 | 62.5 | 66.7 | 67.7 | 64.7 |
| GACGACAAGACCGCG | 57.9 | 57.0 | 54.2 | 61.5 | 59.5 | 57.0 | 64.4 | 62.3 | 60.1 | 67.6 | 67.3 | 65.6 | 68.6 | 69.7 | 68.2 |
| CAGCCTCGTCGCAGC | 60.8 | 60.1 | 58.8 | 64.1 | 62.6 | 61.1 | 67.4 | 65.6 | 63.8 | 70.1 | 70.6 | 68.4 | 72.0 | 73.0 | 70.6 |
| CTCGCGGTCGAAGCG | 61.5 | 60.2 | 57.6 | 64.6 | 62.7 | 60.3 | 67.1 | 65.5 | 63.4 | 70.0 | 70.4 | 68.7 | 70.7 | 72.9 | 71.3 |
| GCGTCGGTCCGGGCT | 64.9 | 64.3 | 62.0 | 67.7 | 67.0 | 64.4 | 70.5 | 70.0 | 67.1 | 73.9 | 75.2 | 71.9 | 74.1 | 77.8 | 74.2 |

| Salt | 69 mM | | | 119 mM | | | 220 mM | | | 621 mM | | | 1020 mM | | |
|---|---|---|---|---|---|---|---|---|---|---|---|---|---|---|---|
| Sequence 5'-…-3' | $T_e$ | $T_{UO}$ | $T_u$ | $T_e$ | $T_{UO}$ | $T_u$ | $T_e$ | $T_{UO}$ | $T_u$ | $T_e$ | $T_{UO}$ | $T_u$ | $T_e$ | $T_{UO}$ | $T_u$ |
| TATGTATATTTTGTAATCAG | 44.4 | 45.0 | 44.9 | 47.7 | 47.6 | 47.8 | 52.6 | 50.7 | 51.1 | 57.6 | 56.0 | 56.8 | 61.2 | 58.6 | 59.6 |
| TTCAAGTTAAACATTCTATC | 45.7 | 47.0 | 45.1 | 49.5 | 49.7 | 48.2 | 53.9 | 52.8 | 51.7 | 59.4 | 58.1 | 57.9 | 61.5 | 60.6 | 60.9 |
| TGATTCTACCTATGTGATTT | 49.1 | 49.6 | 48.9 | 53.5 | 52.4 | 51.9 | 57.4 | 55.5 | 55.3 | 62.3 | 61.0 | 61.3 | 64.4 | 63.7 | 64.2 |
| GAGATTGTTTCCCTTTCAAA | 49.3 | 52.4 | 49.6 | 52.8 | 55.2 | 52.7 | 57.6 | 58.3 | 56.2 | 62.6 | 63.7 | 62.4 | 65.3 | 66.3 | 65.4 |
| ATGCAATGCTACATATTCGC | 55.2 | 55.5 | 56.0 | 59.5 | 58.2 | 58.7 | 62.9 | 61.3 | 61.7 | 67.0 | 66.6 | 67.1 | 68.9 | 69.2 | 69.7 |
| CCACTATACCATCTATGTAC | 51.1 | 50.1 | 51.0 | 54.6 | 52.8 | 53.5 | 58.4 | 55.9 | 56.4 | 62.2 | 61.3 | 61.4 | 64.4 | 63.9 | 63.9 |
| CCATCATTGTGTCTACCTCA | 55.6 | 55.4 | 54.9 | 59.5 | 58.1 | 57.6 | 63.1 | 61.3 | 60.7 | 67.3 | 66.7 | 66.1 | 68.5 | 69.4 | 68.7 |
| CGGGACCAACTAAAGGAAAT | 53.7 | 56.3 | 54.8 | 57.7 | 59.0 | 57.7 | 61.7 | 62.2 | 60.9 | 66.7 | 67.6 | 66.5 | 68.5 | 70.3 | 69.3 |
| TAGTGGCGATTAGATTCTGC | 57.0 | 57.2 | 56.6 | 60.6 | 59.9 | 59.3 | 64.8 | 63.0 | 62.5 | 69.1 | 68.4 | 67.9 | 71.2 | 71.1 | 70.6 |
| AGCTGCAGTGGATGTGAGAA | 59.7 | 60.3 | 60.1 | 63.5 | 63.0 | 62.8 | 67.6 | 66.2 | 65.9 | 71.3 | 71.8 | 71.2 | 73.1 | 74.5 | 73.8 |
| TACTTCCAGTGCTCAGCGTA | 60.3 | 61.8 | 61.6 | 64.4 | 64.6 | 64.2 | 67.7 | 67.8 | 67.2 | 71.6 | 73.3 | 72.3 | 73.6 | 76.0 | 74.8 |
| CAGTGAGACAGCAATGGTCG | 59.8 | 59.8 | 59.4 | 63.5 | 62.5 | 62.1 | 67.0 | 65.6 | 65.2 | 71.1 | 70.9 | 70.5 | 72.5 | 73.5 | 73.1 |
| CGAGCTTATCCCTATCCCTC | 56.0 | 57.3 | 56.7 | 60.2 | 60.0 | 59.4 | 64.1 | 63.2 | 62.4 | 68.5 | 68.6 | 67.6 | 70.3 | 71.3 | 70.1 |
| CGTACTAGCGTTGGTCATGG | 59.6 | 59.3 | 59.3 | 63.1 | 62.0 | 61.8 | 66.6 | 65.1 | 64.7 | 70.5 | 70.4 | 69.8 | 71.1 | 72.9 | 72.2 |
| AAGGCGAGTCAGGCTCAGTG | 64.5 | 63.1 | 63.0 | 67.7 | 65.9 | 65.6 | 71.4 | 69.0 | 68.6 | 75.1 | 74.5 | 73.8 | 76.3 | 77.2 | 76.3 |
| ACCGACGACGCTGATCCGAT | 66.0 | 64.7 | 63.6 | 69.1 | 67.5 | 66.7 | 72.6 | 70.6 | 70.2 | 76.8 | 76.0 | 76.4 | 77.3 | 78.7 | 79.4 |
| AGCAGTCCGCCACACCCTGA | 66.5 | 67.1 | 66.5 | 69.9 | 69.9 | 68.9 | 74.0 | 73.2 | 71.7 | 76.9 | 78.8 | 76.5 | 78.5 | 81.6 | 78.8 |
| CAGCCTCGTTCGCACAGCCC | 67.2 | 67.3 | 66.3 | 70.7 | 70.0 | 68.7 | 74.0 | 73.1 | 71.5 | 77.7 | 78.5 | 76.3 | 78.1 | 81.1 | 78.6 |
| GTGGTGGGCCGTGCGCTCTG | 69.2 | 69.6 | 68.7 | 72.7 | 72.4 | 71.0 | 76.2 | 75.5 | 73.6 | 79.6 | 80.9 | 78.1 | 81.0 | 83.6 | 80.2 |
| GTCCACGCCCGGTGCGACGG | 70.9 | 71.6 | 69.5 | 73.9 | 74.3 | 72.0 | 77.3 | 77.4 | 74.8 | 79.8 | 82.8 | 79.7 | 81.1 | 85.4 | 82.0 |
| GATATAGCAAAATTCTAAGTTAATA | 49.1 | 50.1 | 50.0 | 53.5 | 52.9 | 53.0 | 57.7 | 56.0 | 56.5 | 63.3 | 61.5 | 62.6 | 66.1 | 64.2 | 65.6 |
| ATAACTTTACGTGTGTGACCTATTA | 56.6 | 56.9 | 57.2 | 60.7 | 59.7 | 60.2 | 64.7 | 62.9 | 63.6 | 69.6 | 68.5 | 69.4 | 71.8 | 71.2 | 72.3 |
| GTTCTATACTCTTGAAGTTGATTAC | 52.7 | 53.4 | 52.8 | 56.1 | 56.1 | 55.9 | 60.6 | 59.3 | 59.4 | 66.1 | 64.7 | 65.6 | 67.7 | 67.3 | 68.6 |
| CCCTGCACTTTAACTGAATTGTTTA | 57.4 | 59.1 | 58.5 | 61.4 | 61.9 | 61.3 | 65.6 | 65.1 | 64.6 | 70.1 | 70.6 | 70.2 | 72.5 | 73.4 | 73.0 |
| TAACCATACTGAATACCTTTTGACG | 56.5 | 58.0 | 57.5 | 60.2 | 60.8 | 60.5 | 64.3 | 64.0 | 63.8 | 68.9 | 69.5 | 69.7 | 71.3 | 72.2 | 72.5 |
| TCCACACGGTAGTAAAATTAGGCTT | 59.3 | 60.1 | 59.9 | 63.1 | 62.9 | 62.7 | 67.3 | 66.2 | 65.9 | 71.8 | 71.8 | 71.5 | 73.8 | 74.6 | 74.2 |
| TTCCAAAAGGAGTTATGAGTTGCGA | 59.1 | 60.8 | 59.1 | 63.0 | 63.6 | 62.2 | 67.2 | 66.9 | 65.6 | 71.6 | 72.5 | 71.6 | 73.8 | 75.2 | 74.6 |
| AATATCTCTCATGCGCCAAGCTACA | 62.1 | 62.3 | 62.8 | 65.7 | 65.1 | 65.6 | 70.3 | 68.4 | 68.8 | 75.1 | 74.0 | 74.3 | 76.5 | 76.7 | 77.0 |
| TAGTATATCGCAGCATCATACAGGC | 61.2 | 61.3 | 62.5 | 64.7 | 64.1 | 65.1 | 69.1 | 67.3 | 68.1 | 72.8 | 72.8 | 73.3 | 75.0 | 75.5 | 75.8 |
| TGGATTCTACTCAACCTTAGTCTGG | 59.0 | 59.4 | 59.2 | 63.1 | 62.2 | 62.0 | 67.1 | 65.5 | 65.2 | 71.3 | 71.1 | 70.9 | 73.6 | 73.9 | 73.6 |
| CGGAATCCATGTTACTTCGGCTATC | 60.9 | 61.4 | 60.8 | 64.7 | 64.2 | 63.7 | 68.7 | 67.3 | 67.0 | 73.3 | 72.8 | 72.7 | 74.8 | 75.5 | 75.6 |
| CTGGTCTGGATCTGAGAACTTCAGG | 62.1 | 62.2 | 61.9 | 65.8 | 65.0 | 64.8 | 69.6 | 68.2 | 68.1 | 74.2 | 73.8 | 73.9 | 75.6 | 76.6 | 76.7 |
| ACAGCGAATGGACCTACGTGGCCTT | 68.1 | 68.1 | 68.3 | 72.1 | 70.9 | 71.0 | 76.0 | 74.2 | 74.1 | 79.4 | 79.9 | 79.5 | 81.0 | 82.7 | 82.2 |

| Salt | 69 mM | | | 119 mM | | | 220 mM | | | 621 mM | | | 1020 mM | | |
|---|---|---|---|---|---|---|---|---|---|---|---|---|---|---|---|
| Sequence 5'-…-3' | $T_e$ | $T_{UO}$ | $T_u$ | $T_e$ | $T_{UO}$ | $T_u$ | $T_e$ | $T_{UO}$ | $T_u$ | $T_e$ | $T_{UO}$ | $T_u$ | $T_e$ | $T_{UO}$ | $T_u$ |
| AGCAAGTCGAGCAGGGCCTACGTTT | 68.3 | 68.3 | 68.6 | 72.6 | 71.1 | 71.3 | 76.3 | 74.4 | 74.4 | 80.0 | 80.0 | 79.8 | 81.5 | 82.8 | 82.4 |
| GCGAGCGACAGGTTACTTGGCTGAT | 67.0 | 67.1 | 67.1 | 70.8 | 69.9 | 69.9 | 74.7 | 73.1 | 73.0 | 78.6 | 78.6 | 78.5 | 80.1 | 81.3 | 81.1 |
| AAAGGTGTCGCGGAGAGTCGTGCTG | 69.6 | 68.8 | 68.5 | 73.6 | 71.6 | 71.4 | 77.4 | 74.8 | 74.7 | 81.2 | 80.3 | 80.4 | 82.4 | 83.0 | 83.2 |
| ATGGGTGGGAGCCTCGGTAGCAGCC | 70.7 | 71.7 | 71.8 | 74.5 | 74.6 | 74.1 | 78.2 | 78.0 | 76.8 | 81.6 | 83.8 | 81.4 | 83.4 | 86.6 | 83.6 |
| CAGTGGGCTCCTGGGCGTGCTGGTC | 72.0 | 72.9 | 72.6 | 75.6 | 75.8 | 74.9 | 79.2 | 79.0 | 77.6 | 82.6 | 84.7 | 82.1 | 83.4 | 87.5 | 84.3 |
| GCCAACTCCGTCGCCGTTCGTGCGC | 73.5 | 74.2 | 72.5 | 76.5 | 77.0 | 75.2 | 80.6 | 80.1 | 78.3 | 83.2 | 85.5 | 83.7 | 84.6 | 88.1 | 86.3 |
| ACGGGTCCCCGCACCGCACCGCCAG | 76.8 | 78.3 | 77.0 | 79.9 | 81.2 | 79.3 | 84.0 | 84.5 | 82.0 | 87.1 | 90.2 | 86.7 | 88.3 | 93.0 | 88.9 |
| TTATGTATTAAGTTATATAGTAGTAGTAGT | 50.7 | 51.4 | 53.2 | 55.0 | 54.2 | 56.1 | 59.3 | 57.4 | 59.4 | 65.1 | 63.1 | 65.2 | 66.6 | 65.8 | 68.1 |
| ATTGATATCCTTTTCTATTCATCTTTCATT | 53.5 | 55.7 | 55.5 | 58.3 | 58.6 | 58.9 | 62.3 | 61.9 | 62.8 | 68.6 | 67.5 | 69.6 | 70.4 | 70.3 | 72.9 |
| AAAGTACATCAACATAGAGAATTGCATTTC | 58.3 | 58.7 | 58.7 | 61.9 | 61.5 | 61.9 | 66.1 | 64.7 | 65.5 | 71.3 | 70.3 | 71.8 | 73.2 | 73.0 | 75.0 |
| CTTAAGATATGAGAACTTCAACTAATGTGT | 56.8 | 57.4 | 57.9 | 61.3 | 60.2 | 61.0 | 64.9 | 63.5 | 64.7 | 70.5 | 69.1 | 71.0 | 71.8 | 71.8 | 74.1 |
| CTCAACTTGCGGTAAATAAATCGCTTAATC | 60.9 | 61.1 | 60.6 | 64.8 | 63.8 | 63.6 | 68.7 | 67.0 | 67.1 | 74.4 | 72.5 | 73.2 | 75.5 | 75.2 | 76.2 |
| TATTGAGAACAAGTGTCCGATTAGCAGAAA | 61.3 | 62.9 | 62.3 | 65.4 | 65.8 | 65.4 | 69.6 | 69.1 | 69.0 | 74.8 | 74.7 | 75.2 | 76.4 | 77.5 | 78.3 |
| GTCATACGACTGAGTGCAACATTGTTCAAA | 62.7 | 63.7 | 63.3 | 66.8 | 66.5 | 66.3 | 70.8 | 69.7 | 69.8 | 75.9 | 75.2 | 76.0 | 76.9 | 78.0 | 79.0 |
| AACCTGCAACATGGAGTTTTTGTCTCATGC | 64.5 | 65.7 | 65.6 | 68.3 | 68.6 | 68.6 | 72.5 | 71.8 | 71.9 | 77.7 | 77.5 | 77.8 | 78.7 | 80.3 | 80.6 |
| CCGTGCGGTGTGTACGTTTTATTCATCATA | 63.9 | 65.6 | 65.5 | 68.3 | 68.4 | 68.4 | 71.8 | 71.7 | 71.8 | 76.5 | 77.2 | 77.7 | 77.6 | 80.0 | 80.5 |
| GTTCACGTCCGAAAGCTCGAAAAAGGATAC | 64.3 | 65.3 | 63.9 | 68.2 | 68.1 | 67.0 | 72.1 | 71.3 | 70.7 | 77.1 | 76.8 | 77.1 | 78.7 | 79.5 | 80.3 |
| AGTCTGGTCTGGATCTGAGAACTTCAGGCT | 66.3 | 67.1 | 68.0 | 70.4 | 70.1 | 71.0 | 74.5 | 73.4 | 74.5 | 78.8 | 79.3 | 80.5 | 80.6 | 82.2 | 83.5 |
| TCGGAGAAATCACTGAGCTGCCTGAGAAGA | 66.3 | 67.7 | 67.0 | 70.4 | 70.6 | 70.0 | 74.1 | 73.9 | 73.5 | 79.0 | 79.7 | 79.6 | 80.9 | 82.5 | 82.6 |
| CTTCAACGGATCAGGTAGGACTGTGGTGGG | 67.6 | 68.5 | 68.6 | 71.7 | 71.4 | 71.3 | 74.7 | 74.7 | 74.5 | 78.8 | 80.5 | 79.9 | 80.1 | 83.3 | 82.5 |
| ACGCCCACAGGATTAGGCTGGCCCACATTG | 71.3 | 72.5 | 72.8 | 74.7 | 75.5 | 75.3 | 78.5 | 78.8 | 78.1 | 82.7 | 84.7 | 83.0 | 84.0 | 87.5 | 85.4 |
| GTTATTCCGCAGTCCGATGGCAGCAGGCTC | 70.6 | 71.4 | 71.1 | 74.9 | 74.2 | 73.8 | 78.1 | 77.5 | 76.9 | 82.4 | 83.1 | 82.2 | 84.1 | 85.9 | 84.8 |
| TCAGTAGGCGTGACGCAGAGCTGGCGATGG | 72.2 | 73.5 | 73.6 | 75.7 | 76.3 | 76.3 | 79.3 | 79.6 | 79.3 | 83.3 | 85.3 | 84.6 | 84.6 | 88.1 | 87.1 |
| CGCGCCACGTGTGATCTACAGCCGTTCGGC | 72.7 | 74.8 | 74.3 | 76.3 | 77.5 | 77.1 | 79.5 | 80.7 | 80.2 | 83.4 | 86.3 | 85.6 | 84.5 | 89.0 | 88.2 |
| GACCTGACGTGGACCGCTCCTGGGCGTGGT | 74.4 | 76.3 | 76.0 | 78.4 | 79.3 | 78.6 | 81.5 | 82.6 | 81.7 | 85.2 | 88.4 | 86.9 | 86.4 | 91.2 | 89.5 |
| GCCCCTCCACTGGCCGACGGCAGCAGGCTC | 76.3 | 78.8 | 78.1 | 79.8 | 81.7 | 80.4 | 83.6 | 85.1 | 83.1 | 87.1 | 90.9 | 87.7 | 87.7 | 93.8 | 89.9 |
| CGCCGCTGCCGACTGGAGGAGCGCGGGACG | 77.8 | 80.3 | 79.1 | 81.6 | 83.1 | 81.7 | 84.6 | 86.4 | 84.6 | 87.7 | 92.0 | 89.7 | 88.6 | 94.8 | 92.2 |

**Table S3.** Fraction of different NNBPs in the two sequences. As expected, the fraction of non-degenerate NNBPs (AT/TA, TA/AT, GC/CG, CG/GC) is roughly the half of the fraction of the degenerate ones. Only the fraction of TA/AT is slightly under represented in both sequences.

| **NNBP** | **2.2 kbps (%)** | **6.8 kbps (%)** |
|---|---|---|
| AA/TT | 13.7 | 16.4 |
| AC/TG | 10.4 | 11.2 |
| AG/TC | 10.6 | 10.9 |
| AT/TA | 6.3 | 7.6 |
| CA/GT | 13.8 | 14.1 |
| CC/GG | 14.1 | 9.1 |
| CG/GC | 7.1 | 6.0 |
| GA/CT | 12.5 | 12.8 |
| GC/CG | 7.9 | 6.5 |
| TA/AT | 3.6 | 5.4 |

**Table S4**. Elastic parameters of ssDNA at different salt concentration. *d* is the interphosphate distance for each model, $l_p$ is the persistence length of the WLC and *b* is the Kuhn length of the FJC. The mean values were obtained after averaging over 5 molecules for each salt, except for 25 mM and 100 mM that were averaged over 4 molecules and for 50 mM that were averaged over 3.

| | WLC model<br>$d$ = 0.665 nm | FJC model<br>$d$ = 0.59 nm |
|---|---|---|
| **Salt [NaCl]** | **$l_p$ (nm)** | **$b$ (nm)** |
| 10 mM | 1.14 (0.1) | - |
| 25 mM | 0.93 (0.1) | - |
| 50 mM | 0.88 (0.1) | - |
| 100 mM | - | 1.37 (0.1) |
| 250 mM | - | 1.25 (0.1) |
| 500 mM | - | 1.20 (0.1) |
| 1000 mM | - | 1.15 (0.1) |